\documentclass[
 reprint,
 superscriptaddress,
 bibnotes,
 amsmath,
 amssymb,
 aps,
 prb,
 citeautoscript,
 floatfix,
]{revtex4-2}

\usepackage{graphicx}
\usepackage{dcolumn}
\usepackage{bm}
\usepackage[dvipsnames]{xcolor}
\usepackage[colorlinks=true, citecolor={blue!80!black}, urlcolor={blue!50!black}, linkcolor = {blue!80!black}]{hyperref}
\usepackage{lipsum}  
\usepackage{upgreek}
\usepackage{siunitx}
\usepackage{mathtools}
\usepackage{ulem}
\usepackage{verbatim}

\usepackage{nameref,zref-xr}
\zxrsetup{toltxlabel}

\newcommand{\flux}{$\phi_{\rm ext}$}

\newcommand{\ket}[1]{\left|#1\right\rangle}

\begin{document}

\preprint{APS/123-QED}

\title{Spectroscopy of spin-split Andreev levels in a quantum dot with superconducting leads}

\author{Arno Bargerbos}
\thanks{These two authors contributed equally.}
\affiliation{QuTech and Kavli Institute of Nanoscience, Delft University of Technology, 2600 GA Delft, The Netherlands}

\author{Marta Pita-Vidal}
\thanks{These two authors contributed equally.}
\affiliation{QuTech and Kavli Institute of Nanoscience, Delft University of Technology, 2600 GA Delft, The Netherlands}

\author{Rok Žitko}
\affiliation{Jožef Stefan Institute, Jamova 39, SI-1000 Ljubljana, Slovenia}
\affiliation{Faculty of Mathematics and Physics, University of Ljubljana, Jadranska 19, SI-1000 Ljubljana, Slovenia}

\author{Lukas J. Splitthoff}
\affiliation{QuTech and Kavli Institute of Nanoscience, Delft University of Technology, 2600 GA Delft, The Netherlands}

\author{Lukas Grünhaupt}
\affiliation{QuTech and Kavli Institute of Nanoscience, Delft University of Technology, 2600 GA Delft, The Netherlands}

\author{Jaap J. Wesdorp}
\affiliation{QuTech and Kavli Institute of Nanoscience, Delft University of Technology, 2600 GA Delft, The Netherlands}

\author{Yu Liu}
\affiliation{Center for Quantum Devices, Niels Bohr Institute, University of Copenhagen, 2100 Copenhagen, Denmark}

\author{Leo P. Kouwenhoven}
\affiliation{QuTech and Kavli Institute of Nanoscience, Delft University of Technology, 2600 GA Delft, The Netherlands}

\author{Ramón Aguado}
\affiliation{Instituto de Ciencia de Materiales de Madrid (ICMM),
Consejo Superior de Investigaciones Cientificas (CSIC), Sor Juana Ines de la Cruz 3, 28049 Madrid, Spain}

\author{Christian Kraglund Andersen}
\affiliation{QuTech and Kavli Institute of Nanoscience, Delft University of Technology, 2600 GA Delft, The Netherlands}

\author{Angela Kou}
\affiliation{Department of Physics and Frederick Seitz Materials Research Laboratory,
University of Illinois Urbana-Champaign, Urbana, IL 61801, USA}

\author{Bernard van Heck}
\affiliation{Leiden Institute of Physics, Leiden University, Niels Bohrweg 2, 2333 CA Leiden, The Netherlands}

\date{\today}

\begin{abstract}
We use a hybrid superconductor-semiconductor transmon device to perform spectroscopy of a quantum dot Josephson junction tuned to be in a spin-1/2 ground state with an unpaired quasiparticle. Due to spin-orbit coupling, we resolve two flux-sensitive branches in the transmon spectrum, depending on the spin of the quasi-particle. A finite magnetic field shifts the two branches in energy, favoring one spin state and resulting in the anomalous Josephson effect. We demonstrate the excitation of the direct spin-flip transition using all-electrical control. Manipulation and control of the spin-flip transition enable the future implementation of charging energy protected Andreev spin qubits.
\end{abstract}

\maketitle


In the confined geometry of a Josephson junction, the Andreev reflection of an electron into a hole at a normal-superconducting interface leads to the formation of discrete Andreev bound states (ABS) \cite{Furusaki1990, Beenakker1991, Sauls2018}.
ABS are of fundamental importance from the perspective of mesoscopic superconductivity and are also at the basis of several qubit proposals~\cite{Desposito2001, Zazunov2003a, Chtchelkatchev2003, Padurariu2010}.
In particular, when an ABS is populated by a single quasi-particle, the trapped quasi-particle can serve as the superconducting version of a spin qubit.
In the presence of spin-orbit coupling, the Josephson phase difference $\phi$ may break the degeneracy between the spin states, coupling the spin degree of freedom to the supercurrent across the junction \cite{Chtchelkatchev2003, Beri2008} and allowing for direct integration of spin qubits into superconducting circuits for remote communication, transduction, or hybrid qubit platforms~\cite{Lauk2020, Aguado2020, Spethmann2022}.

Experimental work on InAs/Al nanowire Josephson junctions has shown the presence of the predicted spin-split ABS \cite{Tosi2018, Hays2020, Metzger2021}, culminating in the first demonstration of coherent Andreev spin qubit manipulation \cite{Hays2021}.
These remarkable experiments were operated in a regime in which the spin-1/2 occupation of the junction was an excited manifold and were thus susceptible to qubit leakage via quasiparticle escape or recombination, bringing the junction back into its spin-zero ground state.
Furthermore, the direct manipulation of the spin states proved unfeasible, likely due to the smallness of relevant matrix elements~\cite{Park2017}, requiring complex excitation schemes involving auxiliary levels \cite{Hays2021, Cerrillo2021}.

Recently~\cite{Bargerbos2022}, we showed that by embedding a gate-controlled quantum dot in the InAs/Al Josephson junction it is possible to tune its ground state to be an odd-parity spin-1/2 ground state.
In this doublet phase, the lifetime of the trapped quasi-particle can exceed \SI{1}{ms}, likely benefiting from the large charging energy of the quantum dot suppressing quasiparticle poisoning events.

In this Letter, employing the same transmon techniques as in Ref.~\cite{Bargerbos2022},
we report the detection of the spin splitting of the doublet states of a quantum dot Josephson junction due to the spin-orbit interaction in InAs.
The energy difference between spin states is smaller than the electron temperature, which would make it difficult to detect in transport measurements.
We also demonstrate that the population of the spin-split states, as well as the spin-selective transmon frequencies, can be controlled via modest external magnetic fields smaller than \SI{40}{mT}.
In the presence of magnetic field, we furthermore observe the anomalous Josephson effect: a shift of the minimum of the energy-phase relation to a spin-dependent value $\phi_0$ which is neither $0$ nor $\pi$ \cite{Zazunov2009, Brunetti2013, Szombati2016}.
Finally, we show that the spin states can be directly manipulated by applying microwaves to a bottom gate, via the electric dipole spin resonance (EDSR) \cite{Rashba2003, Flindt2006, Golovach2006, Nowack2007, NadjPerge2010, Berg2013}.
Our experiment is directly comparable to, and inspired by, the theoretical proposal of Ref.~\cite{Padurariu2010}, which we use to model the data, combined with further understanding based on a modified single-impurity Anderson model (SIAM).


\begin{figure}
    \centering
    \includegraphics[scale=1.0]{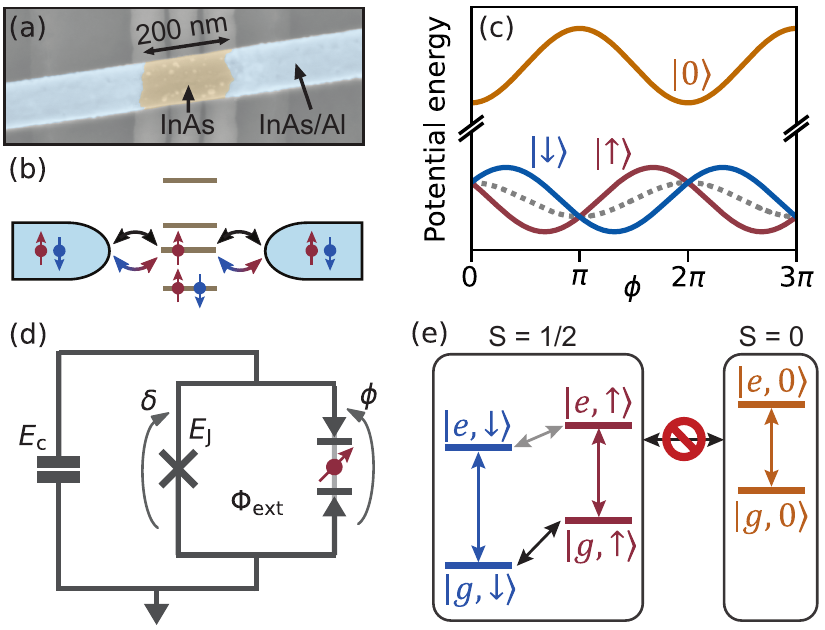}
    \caption{(a) False-colored scanning electron micrograph of the quantum dot junction. (b) Conceptual diagram of the quantum dot Josephson junction with multiple energy levels. Black and blue-to-red arrows denote the spin-conserving and spin-flipping tunnel couplings, respectively. 
    (c) Dependence of the Josephson potential $U$ on the superconducting phase difference between the leads $\phi$, shown for different quantum dot junction states: the singlet state is shown in orange, and the two doublet states are shown in red and blue. The dotted gray line indicates the potential of the doublet states in the absence of the $E_{\rm SO}$ term. (d) Circuit model for a transmon with charging energy $E_{\rm c}$ and a grounded SQUID formed by the parallel combination of a quantum dot junction and a reference Josephson junction with Josephson energy $E_{\rm J}$. $\delta$ denotes the phase difference across the reference junction, and $\Phi_\textrm{ext}$ is the externally applied magnetic flux through the SQUID loop. (e) Level diagram of the joint transmon-quantum dot junction system. The transmon transition frequencies $\ket{g} \leftrightarrow \ket{e}$ (vertical arrows) depend on the state of the quantum dot junction. Coherent microwave transitions between singlet and doublet are forbidden by superselection rules. However, intra-doublet spin-flip transitions are possible, indicated with diagonal arrows.}
    \label{fig:intro}
\end{figure}

At the core of our experiment lies the quantum dot Josephson junction, which is hosted in a nominally \SI{10}{\micro m}-long InAs/Al superconductor-semiconductor nanowire with a \SI{110}{nm}-wide hexagonal core and a \SI{6}{nm}-thick shell covering two facets \cite{Krogstrup2015}.
The quantum dot is electrostatically defined in a $\SI{200}{nm}$-long wet-etched section of InAs using three bottom gate electrodes, and its superconducting leads are formed by the flanking InAs/Al sections [Fig.~\ref{fig:intro}(a)]. The bottom gate voltages can be used to control the occupation of the quantum dot and its coupling to the superconducting electrodes. This results in two possible ground states of the quantum dot junction, which can be either a spin-zero or a spin-1/2 state. We are particularly interested in the latter case [Fig.~\ref{fig:intro}(b)], where the ground state manifold is spanned by the two components $\ket{\downarrow}$ and $\ket{\uparrow}$ of a Kramers doublet, and a minimal model for the potential energy of the quantum dot junction is given by~\cite{Padurariu2010}
\begin{equation}
U(\phi) = E_0\cos\left(\phi\right) - E_{\rm SO}\, \vec{\sigma} \cdot \vec{n}\,\sin\left(\phi\right) +\frac{1}{2} \vec{E}_{\rm Z} \cdot\vec{\sigma}\,.
\label{eq:ESOpotential}
\end{equation}
Here, $\vec{\sigma}$ is the spin operator, $\vec{n}$ is a unit vector along the polarization direction set by the spin-orbit interaction, and $E_{\rm SO}$ and $E_0$ are  the spin-dependent and spin-independent contributions to the Cooper pair tunneling rate across the quantum dot junction.
Note that the term proportional to $E_0$ has a minimum at $\phi=\pi$. This $\pi$-shift originates from the odd occupancy of the junction~\cite{Glazman1989,spivak1991} and distinguishes the Josephson energy from that of a conventional tunnel junction.  Finally, $\vec{E}_{\rm Z}$ is a Zeeman field arising in the presence of an external magnetic field.

The energy scales $E_0$ and $E_{\rm SO}$ can be understood as follows~\cite{Padurariu2010}:
Cooper pair tunneling occurs via a sequence of single-electron co-tunneling processes through the quantum dot energy levels.
The spin-independent component $E_0$ arises from those sequences in which both electrons co-tunnel through the same energy level.
The amplitude for these sequences is the same whether the initial state of the quantum dot junction is $\ket{\downarrow}$ or $\ket{\uparrow}$.
On the other hand, $E_{\rm SO}$ arises from tunneling processes in which one electron co-tunnels through the singly-occupied level, involving a spin rotation, while the second one co-tunnels through a different level.
Since in the presence of spin-orbit coupling the single-electron tunneling amplitudes can be spin-dependent, for these processes the \textit{pair} tunneling amplitude may depend on the spin of the initial state.

The two potential energy branches of the doublet states at $\vec{E}_{\rm Z}=\vec{0}$, $E_{\downarrow, \uparrow} = E_{0} \cos{\phi} \pm  E_{\rm SO} \sin{\phi}$, are sinusoidals with an amplitude of $\sqrt{E_0^2+E_{\rm SO}^2}$ and minima at $\phi_0 = \pi \pm \mathrm{arctan}\left(E_{\rm SO}/E_0\right)$, see Fig.~\ref{fig:intro}(c).
If $E_{\rm SO}=0$, the potential energy reduces to the well-known $\pi$-junction behavior without spin-splitting.
At finite $E_{\rm SO}$, the shift of the minima away from $\phi=0,\pi$ is a precursor~\cite{Padurariu2010} to the anomalous Josephson effect \cite{Zazunov2009, Brunetti2013}; while at $\phi=0$ there will be instantaneous supercurrents on timescales short compared to the spin lifetime, the time-averaged current will be zero due to thermal fluctuations. For completeness, in Fig.~\ref{fig:intro}(c) we also show the potential energy $E_{\rm s}$ corresponding to the lowest-energy singlet state $\ket{0}$, with its familiar minimum at $\phi=0$, as expected for a conventional Josephson junction.

We can derive the occurrence of both $E_0$ and $E_{\rm SO}$ within a minimally-extended SIAM with superconducting leads.
The SIAM is a simple model widely used to understand quantum dot junctions, containing a single energy level coupled to the leads via spin-conserving tunneling events~\cite{Glazman1989, yoshioka2000, Choi2004, Oguri2004, tanaka2007josephson, MartinRodero2011, Karrasch2008, Luitz2012, Kadlecova2019, Meden2019}.
We find that only two extensions to the SIAM are required to generate the spin-splitting term $E_{\rm SO}$: (i) spin-flipping single-electron tunneling between the leads and the energy level~\cite{Danon2009, Padurariu2010, Spethmann2022, Barakov2022} and (ii) inter-lead tunneling,  resulting from integrating out the additional energy levels of the quantum dot.
These results are derived in Sec.~I of the Supplementary Material, together with a validation based on numerical renormalization group calculations \cite{Note3}.
In view of the strong spin-orbit coupling in InAs nanowires~\cite{Fasth2007, Liang2012} and the lateral and transversal confinement on the order of~\SI{100}{nm} \cite{VanDam2006a, Chang2013, Veen2019}, we expect both spin-flipping and spin-conserving tunneling, as well as additional quantum dot levels, to be present in our experimental system.
Note that within this model, the energy $E_0$ in Eq.~\eqref{eq:ESOpotential} may have either sign depending on the relative strength of the two terms.
While both situations may occur at different gate settings in the same device, the tuning procedure to find a doublet ground state relies on the detection of a $\pi$-shift~\cite{Bargerbos2022}.
Thus, our experiment naturally selects the case $E_0>0$, justifying the sign choice in Eq.~\eqref{eq:ESOpotential}.

To experimentally resolve the predicted spin-splitting we follow the method introduced in \cite{Bargerbos2022} and incorporate the quantum dot junction into the superconducting quantum interference device (SQUID) of a transmon circuit [Fig.~\ref{fig:intro}(d)] \cite{Koch2007}.
The different potential energies corresponding to the states $\ket{0}$, $\ket{\downarrow}$, or $\ket{\uparrow}$ give rise to distinct transition frequencies of the transmon circuit [Fig.~\ref{fig:intro}(e)], which can be detected and distinguished via standard circuit quantum electrodynamics techniques \cite{Blais2004, Blais2021}.
We refer to the Supplementary for further details on the device implementation (Sec.~II) \cite{Note3}. 


\begin{figure}[t!]
    \centering
        \includegraphics[scale=1.0]{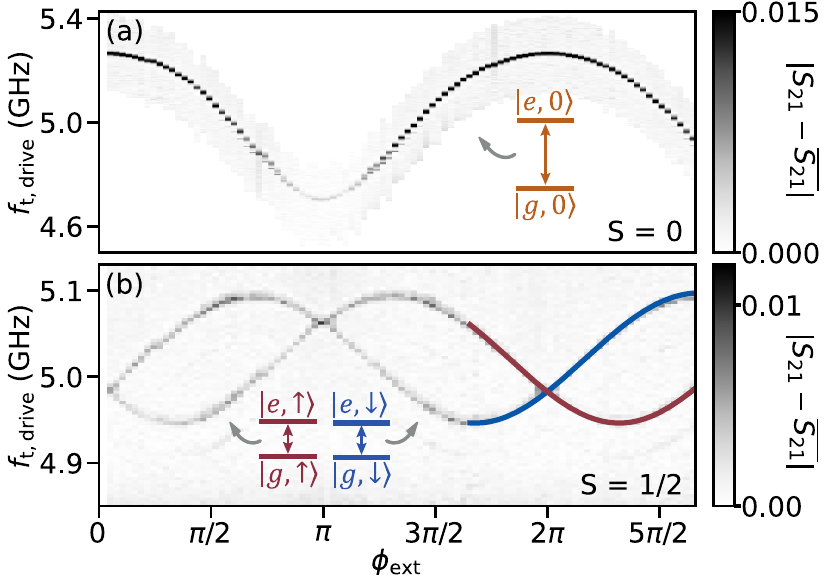}    \caption{Comparison of singlet and spin-split doublet ground states in transmon two-tone spectroscopy. (a) Transmitted microwave signal with varying external flux, \flux, and transmon drive frequency $f_{\rm t, drive}$ for the quantum dot junction in the singlet state. (b) Same as (a) for the junction in the doublet state, at gate configuration A. The two ground states occur for two different gate voltage settings, as detailed in Refs.~\cite{Bargerbos2022,Note3}. The comparison reveals the spin-splitting of the doublet state. Solid lines show fits to a transmon circuit model containing Eq.~\eqref{eq:ESOpotential} \cite{Note3}.}   
    \label{fig:transmon-spec}
\end{figure}

In order to study the system in the regime of interest, we tune the quantum dot junction to a spin-1/2 ground state, as detailed in Ref.~\cite{Note3}, where we refer to this gate setpoint as configuration A. This is followed by a two-tone spectroscopy measurement for which we apply both tones through the feedline and detect the transmon transition frequency as a function of the applied flux $\Phi_\textrm{ext}$.
We find that the flux is tuned with a small in-plane external magnetic field applied perpendicular to the wire~\cite{Wesdorp2022}, requiring a \SI{1.8}{mT} field for adding one flux quantum through the SQUID.
We note that, since the reference junction is tuned to have a Josephson energy $E_{\rm J}/h=12.5$~GHz, much higher than that of the quantum dot junction, the phase difference across the latter is well approximated by $\phi_\textrm{ext}=2e\Phi_\textrm{ext}/\hbar$.

In Fig.~\ref{fig:transmon-spec}(a), we show the typical flux dispersion observed in two-tone spectroscopy when the gate voltages are such that the ground state is a singlet, with the maximum frequency occurring at $\phi_\textrm{ext}=0$.
In fact, this measurement serves as a calibration of the applied flux, which is assumed to be an integer multiple of the flux quantum when the transmon frequency is maximal.
In contrast, when the ground state is electrostatically set to be a doublet, the transition spectrum displays two shifted frequency branches, with maxima at $\phi_{\rm ext}  = \phi_0 \neq 0,\pi$ [Fig.~\ref{fig:transmon-spec}(b)].
The measured spectrum is in good agreement with that predicted by a transmon circuit model with the potential energy of Eq.~\eqref{eq:ESOpotential}, with $E_0/h = \SI{190}{MHz}$ and $E_{\rm SO}/h = \SI{300}{MHz}$.
The latter corresponds to a temperature scale of \SI{14}{mK} \cite{Note3}, indicating that transmon-based spectroscopy can experimentally resolve the spin-orbit splitting of the doublet state well below the thermal broadening that typically limits tunneling spectroscopy experiments. Furthermore, the simultaneous observation of both transmon branches is indicative of a large thermal occupation of the excited spin state, which prevents the splitting from being observable in switching current measurements as the two contributions to the current cancel out. 

We note that tuning the junction to a spin-1/2 ground state is not a sufficient condition to observe the spin-splitting in the transition spectrum.
By tuning the quantum dot to different resonances, corresponding to a quasi-particle trapped to different levels of the quantum dot, we frequently find instances of doublets without the predicted splitting, such as the one studied in detail in Ref.~\cite{Bargerbos2022}.
There are also doublet states that show a small, MHz-size spin-splitting comparable to the transmon linewidth, as well as doublet states with even larger splittings than shown in Fig.~\ref{fig:transmon-spec}(b).
This range of behaviours is shown in Sec.~IV~A of the Supplementary Materials \cite{Note3}.
We attribute the variability of spin-splittings encountered across the device to mesoscopic fluctuations~\cite{Padurariu2010}, due to factors outside of our experimental control, such as disorder and confinement effects on the quantum dot wave functions.


\begin{figure}[t!]
    \centering
    \includegraphics[scale=1.0]{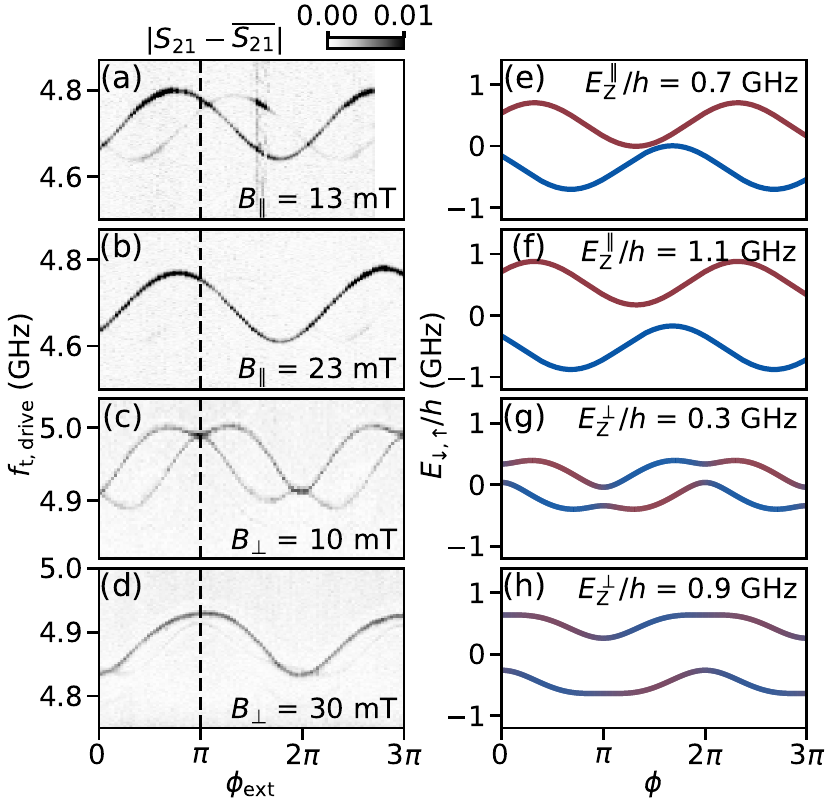} \caption{Magnetic field dependence of the doublet states for gate configuration A.
    (a-d) Two-tone spectroscopy versus~\flux~showing the transmon transitions for a magnetic field applied either parallel (a,b) or perpendicular (c,d) to the inferred spin-orbit direction $\vec{n}$. The vertical dashed lines indicate \flux~$=\pi$ for each panel.
    (e-f) Numerically calculated $\phi$~dependence of the quantum dot junction's Josephson potential for the two doublet states, obtained by diagonalizing the model in Eq.~\eqref{eq:ESOpotential}, in the presence of parallel (e-f) and perpendicular (g-h) Zeeman field. Blue and red colors denote $\ket{\downarrow}$ and $\ket{\uparrow}$ spin polarization, respectively, with a blend of the two indicating mixing of the states. Note that panels (e-h) are not fits to the data of (a-d). Instead, together with the contribution from the reference junction, constitute the potentials that determine the transmon energy levels and serve to build a qualitative understanding (see text). }
    \label{fig:field}
\end{figure}

The transition spectrum is affected by magnetic fields through the Zeeman interaction and depends sensitively on the direction of the applied field with respect to the spin-orbit direction $\vec{n}$, as seen in Fig.~\ref{fig:field}, which shows the behaviour for two limiting cases denoted as the $B_\parallel$ and $B_\perp$ directions. The evolution for intermediate directions and the procedure used to infer the spin-orbit direction are discussed in Supplementary Materials Sec.~IV~C \cite{Note3}.
The flux dispersion of the transition frequencies is only weakly affected by increasing $B_\parallel$ \footnote{We attribute the field-dependent change in flux dispersion to the renormalization of the impurity $g$-factor by coupling to the leads, known as the impurity Knight shift. A detailed understanding of this effect in the context of the superconducting quantum dot will require further investigation.}. Moreover, one of the two spin branches gradually disappears [Fig.~\ref{fig:field}(a)] until at $B_\parallel\gtrsim 23$ mT only a single spectroscopic line remains visible [Fig.~\ref{fig:field}(b)].
In this regime, the minimum transition frequency of the single-valued dispersion remains away from either $0$ or $\pi$, and thus the junction exhibits the anomalous Josephson effect \cite{Zazunov2009, Brunetti2013, Szombati2016}.
In contrast, increasing the magnetic field along the $B_\perp$ direction appears to couple the two spectroscopic lines, leading to branches with two minima per flux period [Fig.~\ref{fig:field}(c)].
At even higher fields this behaviour is lifted, and once-more only one of the two transmon branches persists [Fig.~\ref{fig:field}(d)].
In this case, however, the $\phi_0$-offset has strongly decreased. 

The observed behaviour can be qualitatively understood from Eq.~\eqref{eq:ESOpotential} by considering the cases in which the Zeeman field is respectively parallel or perpendicular to the spin-orbit field.
A parallel Zeeman field $E_{\rm Z}^{\parallel}$ separates the doublet potentials in energy without distorting their phase dependence [Fig.~\ref{fig:field}(e-f)].
As the energy separation increases, the thermal population of the higher-energy state should decrease and, with it, so should the visibility of the corresponding transmon frequency branch.
As the transmon frequency is insensitive to overall shifts in the energy-phase relation, it remains largely unaffected by the $\phi$-independent field-induced energy shift.
A Zeeman term $E_{\rm Z}^{\perp}$ perpendicular to the spin-orbit direction instead couples the two states and opens up an avoided crossing in the Josephson potential energy [Fig.~\ref{fig:field}(g)].
This results in the peculiar flux dependence seen in spectroscopy for moderate fields [Fig.~\ref{fig:field}(c)].
Finally, when $E_{\rm Z}^{\perp}$ becomes much larger than $E_{\rm SO}$, the doublet states instead polarize along the applied field direction [Fig.~\ref{fig:field}(h)], suppressing the $\phi_0$-offset \cite{Zazunov2009, Szombati2016}. 

We note that the $B_\parallel$ and $B_\perp$ field directions do not appear to be directly related to the orientation of the nanowire, as $B_\parallel$ points 13 degrees away from the nanowire axis \cite{Note3}.
This behavior is different from that of long single-gated semiconducting Josephson junctions, where the direction of the spin-orbit interaction, $\vec{n}$, is found to be perpendicular to the nanowire axis~\cite{Tosi2018, Strambini2020}.
Again, we attribute this to mesoscopic fluctuations, supported by the finding that the $B_\parallel$ and $B_\perp$ directions are unique to each region in gate space~\cite{Han2022, Note3}.


\begin{figure}[t!]
    \centering
        \includegraphics[scale=1.0]{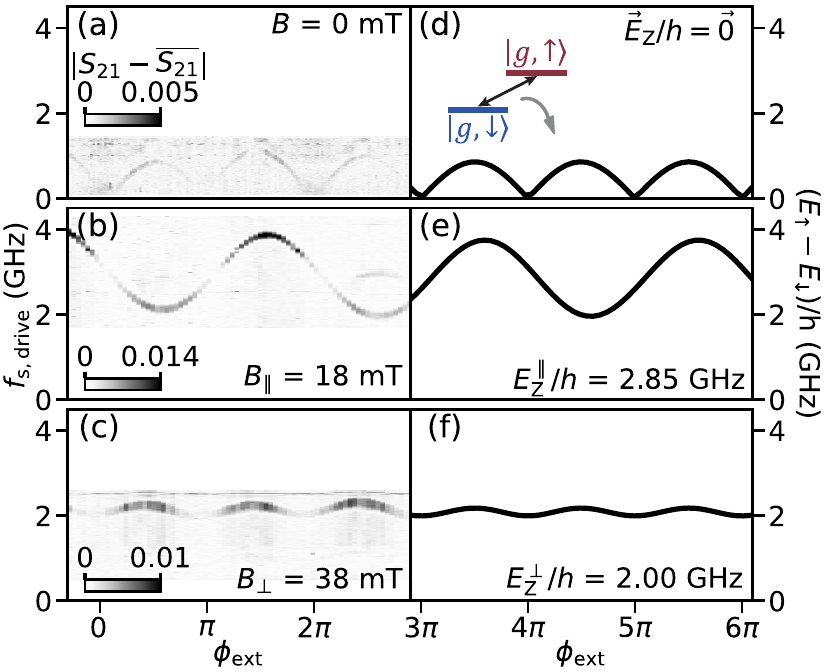}    \caption{Direct spin-flip spectroscopy for gate configuration B.
        (a-c) Measured flux dependence of the direct $\ket{g,\downarrow} \leftrightarrow \ket{g,\uparrow}$ transition frequency for no (a), parallel (b), and perpendicular (c) magnetic field applied relative to the spin-orbit direction. 
        (d-f) Numerically calculated flux dependence of the $\ket{g,\downarrow} \leftrightarrow \ket{g,\uparrow}$ transition frequency for no (d), parallel (e), and perpendicular (f) Zeeman fields applied relative to the spin-orbit direction.
        }
    \label{fig:spin-flip}
\end{figure}


In order to use the doublet states as a superconducting (Andreev) spin qubit \cite{Chtchelkatchev2003, Padurariu2010}, the ability to drive transitions between the doublet states is crucial.
Encouragingly, a recent work by our group indicates that spin-flip transitions of Andreev bound states are possible in the presence of an external magnetic field~\cite{Wesdorp2022}.
Further motivated by all-electrical microwave excitation of electron spins in quantum dots via EDSR~\cite{Nowack2007,NadjPerge2010}, we apply a microwave tone directly to the central bottom gate electrode to excite the doublet states.
For this measurement, we tune the transmon transition frequency close to the readout resonator mode, considerably enhancing its dispersive shift so that we can detect the state of the quantum dot junction while keeping the transmon in its ground state. In addition, we tune away from the gate configuration investigated so far (configuration A) to a parameter regime with a larger spin-splitting $E_{\rm SO}/h=\SI{560}{MHz}$ (configuration B) to maximize the visibility of the doublet splitting \cite{Note3}. 

Applying a microwave drive to the central gate electrode, we find that a low-frequency transition of up to \SI{1}{GHz} can be detected for a vanishing applied magnetic field~\footnote{The tuning of the external flux requires a small magnetic field, which for the data in Fig.~\ref{fig:spin-flip}(a) ranges from \SI{-1.5}{mT}~to~\SI{0.7}{mT}.}, as also shown in Ref.~\cite{Metzger2021}, although the visibility is poor, see Fig.~\ref{fig:spin-flip}(a). This is potentially due to the large thermal population of the excited state, reducing the achievable change in dispersive shift, as well as due to the lack of magnetic field, reducing the efficacy of EDSR \cite{Flindt2006, Golovach2006}.
At elevated $B_\parallel$ the transition frequency rises and becomes well-resolved, with the transition now visible over almost the entire flux range [Fig.~\ref{fig:spin-flip}(b)].
For an applied perpendicular field the transition frequency increases more slowly, and its flux periodicity is half that of the transition in the parallel field direction [Fig.~\ref{fig:spin-flip}(c)]. 

The observed behaviour is consistent with the expected transitions between the doublet states [Fig.~\ref{fig:spin-flip}(d-f)], with the period-halving in perpendicular field being a result of the avoided crossings between the spin branches [c.f. Fig.~\ref{fig:field}(g)].
The comparison to the model furthermore allows us to estimate the effective Landé g-factor in the parallel and perpendicular directions, $g_\parallel = 11$ and $g_\perp = 3.8$ respectively.
We note that the value, as well as the anisotropy, of the g-factor depend strongly on the applied gate voltages (see Supplementary Materials Sec.~IV~B), likely tied to an interplay of spin-orbit coupling and confinement, beyond the scope of the model considered here \cite{Kiselev1998, Csonka2008, Schroer2011, Winkler2017}.

To conclude, our microwave measurements have revealed the rich spin structure of energy levels in a quantum dot Josephson junction, as well as the occurrence of the anomalous Josephson effect. These findings are promising for applications in superconducting spintronics \cite{Linder2015, Shukrinov2022}.
The ability to drive direct spin-flip transitions between the doublet states has strong implications for the nascent field of Andreev spin qubits \cite{Chtchelkatchev2003, Padurariu2010}, since so far their coherent manipulation relied on fine-tuned Raman transitions through auxiliary levels \cite{Cerrillo2021, Hays2021}.
Having direct, all-electrical access to these transitions promises simpler and faster qubit manipulation \cite{NadjPerge2010, Berg2013}.
Furthermore, the polarization of the doublet states at elevated magnetic fields eliminates the unwanted excited state population observed in previous investigations~\cite{Hays2020, Hays2021}.
In fact, we have been able to implement an Andreev spin qubits exploiting these features of our setup, which will be reported in a forthcoming work.
Finally, the demonstrated tunability of the transition frequency, enabled by both flux and magnetic field, is a necessary ingredient for scalable networks of such qubits \cite{Spethmann2022}.


\begin{acknowledgments}
We acknowledge fruitful discussion with Yuli Nazarov and Valla Fatemi. We further thank Peter Krogstrup for guidance in the material growth. This research was inspired by prior work by co-author J.J.W. where the spin-flip transition in an InAs/Al nanowire weak-link was directly observed in spectroscopy under the application of a magnetic field~\cite{Wesdorp2022}. This research is co-funded by the allowance for Top consortia for Knowledge and Innovation (TKI’s) from the Dutch Ministry of Economic Affairs, research project {\it Scalable circuits of Majorana qubits with topological protection} (i39, SCMQ) with project number 14SCMQ02, from the Dutch Research Council (NWO), and the Microsoft Quantum initiative. R. \v{Z}. acknowledges the support of the Slovenian Research agency (ARRS) under P1-0416 and J1-3008. R. A. acknowledges support from the Spanish Ministry of Science and Innovation through Grant PGC2018-097018-B-I00 and from the CSIC Research Platform on Quantum Technologies PTI-001. CKA additionally acknowledges support from the Dutch Research Council (NWO). B.v.H. was partially supported by the Dutch Research Council (NWO).\\
\end{acknowledgments}

\section*{Data availability}
The data and analysis code that support the findings of this study will be made available at 4TU.ResearchData.\\

\section*{Author contributions}

A.B., M.P.V., A.K. and B.v.H. conceived the experiment.
Y.L. developed and provided the nanowire materials.
A.B., M.P.V., L.J.S., L.G. and J.J.W prepared the experimental setup and data acquisition tools.
L.J.S. deposited the nanowires.
A.B. and M.P.V. designed and fabricated the device, performed the measurements and analysed the data, with continuous feedback from L.J.S., L.G., J.J.W, C.K.A, A.K. and B.v.H.
R.A., B.v.H. and R.\v{Z}. provided theory support during and after the measurements and formulated the theoretical framework to analyze the experiment.
R.\v{Z}. performed the analytical and numerical calculations.
A.B., M.P.V. and B.v.H. wrote the code to compute the circuit energy levels and extract experimental parameters.
L.P.K., R.A., C.K.A., A.K. and B.v.H. supervised the work.
A.B., M.P.V., R.\v{Z}. and B.v.H. wrote the manuscript with feedback from all authors.

\footnotetext[3]{See Supplemental Material, which contains further details about theoretical modeling, device fabrication, experimental setup, device tune-up,  extended data on the spin splitting versus gate voltage and magnetic field angle, as well as details of the spin-flip spectroscopy method. It includes Refs.~\cite{Karrasch2009, affleck2000,PhysRevB.62.6687,vecino2003,oguri2004josephson,meng2009, Bargerbos2020, Kringhoj2020b, Kringhoj2018, Spanton2017, Hart2019, Serniak2019, Uilhoorn2021}.}
\bibliography{ms.bib}

\end{document}


\beginsupplement

\title{Supplementary information for ``Spectroscopy of spin-split Andreev levels in a quantum dot with superconducting leads''}

\author{Arno Bargerbos}
\thanks{These two authors contributed equally.}
\affiliation{QuTech and Kavli Institute of Nanoscience, Delft University of Technology, 2600 GA Delft, The Netherlands}

\author{Marta Pita-Vidal}
\thanks{These two authors contributed equally.}
\affiliation{QuTech and Kavli Institute of Nanoscience, Delft University of Technology, 2600 GA Delft, The Netherlands}

\author{Rok Žitko}
\affiliation{Jožef Stefan Institute, Jamova 39, SI-1000 Ljubljana, Slovenia}
\affiliation{Faculty of Mathematics and Physics, University of Ljubljana, Jadranska 19, SI-1000 Ljubljana, Slovenia}

\author{Lukas J. Splitthoff}
\affiliation{QuTech and Kavli Institute of Nanoscience, Delft University of Technology, 2600 GA Delft, The Netherlands}

\author{Lukas Grünhaupt}
\affiliation{QuTech and Kavli Institute of Nanoscience, Delft University of Technology, 2600 GA Delft, The Netherlands}

\author{Jaap J. Wesdorp}
\affiliation{QuTech and Kavli Institute of Nanoscience, Delft University of Technology, 2600 GA Delft, The Netherlands}

\author{Yu Liu}
\affiliation{Center for Quantum Devices, Niels Bohr Institute, University of Copenhagen, 2100 Copenhagen, Denmark}

\author{Leo P. Kouwenhoven}
\affiliation{QuTech and Kavli Institute of Nanoscience, Delft University of Technology, 2600 GA Delft, The Netherlands}

\author{Ramón Aguado}
\affiliation{Instituto de Ciencia de Materiales de Madrid (ICMM),
Consejo Superior de Investigaciones Cientificas (CSIC), Sor Juana Ines de la Cruz 3, 28049 Madrid, Spain}

\author{Christian Kraglund Andersen}
\affiliation{QuTech and Kavli Institute of Nanoscience, Delft University of Technology, 2600 GA Delft, The Netherlands}

\author{Angela Kou}
\affiliation{Department of Physics and Frederick Seitz Materials Research Laboratory,
University of Illinois Urbana-Champaign, Urbana, IL 61801, USA}

\author{Bernard van Heck}
\affiliation{Leiden Institute of Physics, Leiden University, Niels Bohrweg 2, 2333 CA Leiden, The Netherlands}

\date{\today}

\maketitle

\tableofcontents

\vspace{2 cm}

\section{Theory}
\label{Ss:theory}


\newcommand{\beq}[1]{\begin{equation} #1 \end{equation}}
\newcommand{\corr}[1]{\langle\langle #1 \rangle\rangle}
\newcommand{\DD}{\boldsymbol{\Delta}}
\newcommand{\GG}{\mathbf{G}}
\newcommand{\Id}{\mathcal{I}}
\newcommand{\HH}{\mathbf{H}}
\newcommand{\VV}{\mathbf{V}}
\newcommand{\FF}{\mathbf{F}}
\newcommand{\up}{\uparrow}
\renewcommand{\do}{\downarrow}
\renewcommand{\tt}{\tilde{t}}
\newcommand{\XX}{\mathbf{U}}

\begin{figure}[t!]
    \centering
        \includegraphics[scale=1.0]{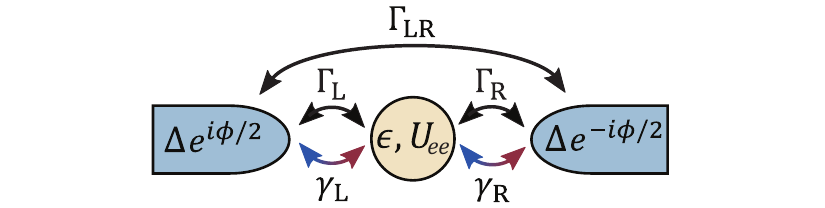}    \caption{  Model diagram of the quantum dot junction. Two $s$-wave superconductors are connected via tunnelling rates to a single level of a quantum dot. $\Gamma_{\rm L, R}$ and $\gamma_{\rm L, R}$ denote, respectively, the spin-conserving and spin-flipping tunneling rates between the superconducting leads and quantum dot. $\Gamma_{\rm LR}$ denotes a spin-conserving effective tunneling rate between the superconducting leads via all remaining energy levels, with 
        $\Gamma_{\rm LR}=\pi \rho |t|^2$.
        }
    \label{fig:model}
\end{figure}

To solidify our understanding of the results and of the mechanisms that govern the size of the spin splitting, we set up a minimal model that is able to reproduce the qualitative features observed experimentally.
Our starting point is an extension of the single-impurity Anderson model (SIAM) for a quantum dot (QD) attached to two superconducting leads \cite{Meden2019,Bargerbos2022}, see Fig.~\ref{fig:model}.
Compared to the standard SIAM, our model also contains spin-flip tunneling between the impurity and the leads due to the presence of spin-orbit coupling, as well as an additional direct tunneling term between the leads. The non-interacting part of the Hamiltonian is
\beq{
\label{H}
\begin{split}
H_0
&= \sum_\sigma \epsilon d^\dag_\sigma d_\sigma 
    + E_x S_x + E_y S_y + E_z S_z \\
&+ \sum_{i,k\sigma} \epsilon_k c^\dag_{i,k\sigma} c_{i,k\sigma}  
+\sum_{i,k} \Delta_i \left(e^{i\phi_i} c^\dag_{i,k\uparrow}  c^\dag_{i,k\downarrow} + \text{H.c.} \right) \\
&+\sum_{i,k\sigma} \left( V_{i,k} c^\dag_{i,k\sigma} d_\sigma + \text{H.c.} \right)
+ \sum_{i,k\sigma} \left( i W_{i,k} c^\dag_{i,k\sigma} d_{\bar{\sigma}} + \text{H.c.}\right) \\
&+ \sum_{k,k',\sigma} \left( t c^\dag_{{\rm L},k\sigma} c_{{\rm R},k'\sigma} + \text{H.c.} \right).
\end{split}
}
The first line describes the QD level $\epsilon$ closest to the Fermi level (the ``resonant'' level), subject to an external magnetic field $\vec{E}_{\rm Z}$ with the components $E_x$, $E_y$ and $E_z$ expressed in units of energy (i.e., as Zeeman energy contributions).
The operator $d^\dag_\sigma$ is the creation operator for an electron in the resonant level, and $S_x$, $S_y$, $S_z$ are impurity spin operators.
The second line describes two superconductors with the dispersion relation $\epsilon_k$ and order parameters $\Delta_i \exp(i\phi_i)$.
The operator $c^\dag_{i,k\sigma}$ is the creation operator for an electron in the left ($i={\rm L}$) or right ($i={\rm R}$) superconductor, in level $k$ and with spin $\sigma$.
The third line describes the QD-superconductor hybridisation; we include both spin-preserving and spin-flipping processes with amplitudes $V_{i,k}$ and $W_{i,k}$, respectively.
The notation $\bar{\sigma}$ denotes spin inversion, $\bar{\uparrow} = \downarrow$, $\bar{\downarrow} = \uparrow$.
Alternatively, we may characterize the tunnel barriers via tunneling rates $\Gamma_{\rm L}=\pi \rho |V_{{\rm L},k_F}|^2$ or $\Gamma_{\rm R}=\pi \rho |V_{{\rm R},k_F}|^2$ for spin-preserving processes, and $\gamma_{\rm L}=\pi \rho |W_{{\rm L},k_F}|^2$ or $\gamma_{\rm R}=\pi \rho |W_{{\rm R},k_F}|^2$ for spin-flip tunneling.
Here $\rho$ is the normal-state density of states and we took the matrix elements at the Fermi level, hence at $k=k_F$. Finally, the last line accounts for the presence of all other (``non-resonant'') levels in the QD: the electron can also cotunnel through the QD via those high-lying levels, which provides  another conduction pathway through the dot. Formally, we may consider this term to arise from integrating out all other levels in the QD, so that
\beq{
t = \sum_{l,k,k'} \frac{V^*_{{\rm L},k;l} V_{{\rm R},k';l}}{\Delta \epsilon_l},
}
where we sum over all ``non-resonant'' levels, $V_{{\rm L/R},k;l}$ are the corresponding tunneling amplitude, while $\Delta \epsilon_l$ are the energy levels. (For simplicity, we are disregarding interactions and spin-flip processes.)
The inter-lead hopping term makes the model resemble those for a QD embedded in a nanoscopic Aharonov-Bohm ring \cite{karrasch2009}. The model breaks down if the level spacing is too small (less than the scale of $\Gamma_{\rm L/R}$): in that case one should use a multi-orbital Anderson impurity model instead.

In addition to this last term, we could also include the spin-flip tunneling through high-lying levels, however this brings about no new qualitative effect.
As we will show, for what follows, the important element is that the ratio of spin-flip to spin-preserving tunneling rate is different for the resonant level and for the aggregate tunneling rate through all remaining non-resonant levels.
This generic situation is expected to hold in most circumstances due to mesoscopic variability of tunneling matrix elements for different QD levels.
The hopping elements, $V_{i, k}$, $W_{i,k}$ and $t$, are in general complex-valued (``directional''): if we reverse the electron flow direction, the amplitude needs to be complex conjugated.

The interacting part of the Hamiltonian is standard:
\beq{
H_\mathrm{int} = U_{ee} n_\uparrow n_\downarrow,
}
where $U_{ee}$ is the electron-electron repulsion on the QD and $n_\sigma=d^\dag_\sigma d_\sigma$ is the occupancy operator.

The model is analytically tractable in the regime $\Delta_{\rm L}, \Delta_{\rm R} \gg U_{ee}$ (the ``superconducting atomic limit'' \cite{affleck2000,PhysRevB.62.6687,vecino2003,oguri2004josephson,tanaka2007josephson,karrasch2008,meng2009}) and in the regime $\Gamma_{\rm L}, \Gamma_{\rm R} \ll U_{ee}$ (the perturbative limit). We find that it is particularly instructive to integrate out the superconducting electrons and compute the hybridization matrix for this Hamiltonian. This leads to relatively simple closed-form expressions that can be used to construct a highly simplified model. Such a model nevertheless seems to be sufficient to account for the full range of the observed behaviors.
The analytic calculations may be verified with explicit calculations using the numerical renormalization group (NRG) techniques, probing their validity over a wide range of parameters.

\subsection{Analytics}\label{Sss:analytics}

We work in the $4\times4$ Nambu representation with
\beq{
d^\dag_a = \begin{pmatrix}
d^\dag_\uparrow & d_\downarrow & d^\dag_\downarrow & d_\uparrow
\end{pmatrix}.
}
Similar notation is used for the $c_{k}$ operators in the superconducting leads.
We will use indexes from the beginning of the alphabet, $a,b,\ldots$,
to denote the Nambu-space index $1,2,3,4$. We define the Green's function (GF) matrices $G_{ab}(z) = \corr{d_a ; d_b^\dag}_z$ and $F_{ab,ik}(z) = \corr{d_a;c_{ik,b}^\dag}_z$; here $\corr{A;B}_z$ denotes the Laplace transform of the Green's function,
$\corr{A;B}_z = \int_0^\infty e^{izt} \corr{A;B}_t \mathrm{d}t$, where
$\corr{A;B}_t = -i\Theta(t) \langle \left\{ A(t), B(0) \right\} \rangle$ is the retarded GF.
In this notation, the equations of motion (EOMs) take the form
$z \corr{A;B}_z = \langle \left\{ A,B \right\} \rangle + \corr{ A; [H,B] }_z$.
The EOM for the QD take the form
\beq{
\label{eqS5}
z \GG(z) = \Id + \GG(z) \HH_0 + \sum_{i,k} \FF_{ik}(z) \VV_{ik} + \XX(z).
}
Here the argument $z=\omega+i\delta$ is complex frequency, $\Id$ is the identity matrix, $\HH_0$ corresponds to the Nambu matrix representation of the
non-interacting part of the impurity Hamiltonian:
\beq{
\HH_0=\begin{pmatrix}
\epsilon +E_z/2 & 0 & (E_x-iE_y)/2 & 0\\
0 & -\epsilon+E_z/2 & 0 & -(E_x-iE_y)/2 \\
(E_x+iE_y)/2 & 0 & \epsilon-E_z/2 & 0 \\
0 & -(E_x+iE_y)/2 & 0 & -\epsilon - E_z/2
\end{pmatrix},
}
$\VV_{ik}$ contains the hopping matrix elements:
\beq{
\VV_{ik} = \begin{pmatrix}
V_{i,k} & 0 & i W_{i,k} & 0 \\
0 & -V^*_{i,k} & 0 & i W_{i,k}^* \\
i W_{i,k} & 0 & V_{i,k} & 0 \\
0 & i W_{i,k}^* & 0 & -V_{i,k}^*
\end{pmatrix},
}
and $\XX(z)$ contains the contributions of the interacting part of the Hamiltonian, $H_\mathrm{int}$. The EOMs for mixed GFs $F_{ik}$ (dropping the frequency argument $z$ in GFs for clarity) may be written as
\newcommand{\FFF}{\mathcal{F}}
\newcommand{\TT}{\mathbf{T}}
\beq{
\begin{split}
\FF_{{\rm L}k}
\left[
z \Id -
\HH_{{\rm L}k}
\right]
&=
\GG \VV^\dag_{{\rm L}k} + \FFF_{\rm R} \TT^*, \\
\FF_{{\rm R}k}
\left[
z \Id -
\HH_{{\rm R}k}
\right]
&=
\GG \VV^\dag_{{\rm L}k} + \FFF_{\rm L} \TT,
\end{split}
}
with
\beq{
\HH_{ik} = \begin{pmatrix}
\epsilon_k & e^{+i\phi_i} \Delta_i & 0 & 0 \\
e^{-i\phi_i} \Delta_i & -\epsilon_k & 0 & 0 \\
0 & 0 & \epsilon_k & -e^{i\phi_1}\Delta_1 \\
0 & 0 & -e^{-i\phi_1}\Delta_1 & -\epsilon_k
\end{pmatrix},
}
\begin{equation}
\TT = \begin{pmatrix}
t & 0 & 0 & 0 \\
0 & -t^* & 0 & 0 \\
0 & 0 & t & 0 \\
0 & 0 & 0 & -t^*
\end{pmatrix},
\end{equation}
and
\beq{
\FFF_i = \sum_k \FF_{ik}.
}

Using the lead propagator $\GG_{ik} = \left[ z \Id - \HH_{ik} \right]^{-1} $,
this may be rewritten as 
\beq{
\begin{split}
\FF_{{\rm L}k} &= \GG \VV_{{\rm L}k}^\dag \GG_{{\rm L}k} + \FFF_{\rm R} \TT^* \GG_{{\rm L}k},\\
\FF_{{\rm R}k} &= \GG \VV_{{\rm R}k}^\dag \GG_{{\rm R}k} + \FFF_{\rm L} \TT \GG_{{\rm R}k}.\\
\end{split}
}
We now assume that $V_{ik}$ and $W_{ik}$ do not depend on $k$, i.e., $V_{ik} \equiv
V_i$ and $W_{ik} \equiv W_i$; this is a reasonable approximation for the relevant
bath levels in the vicinity of the Fermi level.
We define $\GG_i = \sum_k \GG_{ik}$ and we sum the EOMs over $k$:
\beq{
\begin{split}
\FFF_{\rm L} &= \GG \VV_{\rm L}^\dag \GG_{\rm L} + \FFF_{\rm R} \TT^* \GG_{\rm L},\\
\FFF_{\rm R} &= \GG \VV_{\rm R}^\dag \GG_{\rm R} + \FFF_{\rm L} \TT \GG_{\rm R}.\\
\end{split}
}
or
\beq{
\label{eqpr}
\begin{pmatrix}
\FFF_{\rm L} & \FFF_{\rm R} 
\end{pmatrix}
\begin{pmatrix}
\Id & -\TT \GG_{\rm R} \\
-\TT^* \GG_{\rm L} & \Id
\end{pmatrix}
=
\GG \begin{pmatrix}
\VV_{\rm L}^\dag \GG_{\rm L} & \VV_{\rm R}^\dag \GG_{\rm R} 
\end{pmatrix}.
}
The important observation here is that $\FFF_i$ are proportional to $\GG$. This means
that the third term in the EOM \eqref{eqS5} can be written as
\beq{
\sum_{ik} \FF_{ik}(z) \VV_{ik} = \GG(z) \DD(z),
}
where $\DD(z)$ is the hybridization matrix which describes the renormalization of the QD level due to electron excursions in the superconducting leads.
Eq.~\eqref{eqpr} can be solved for each $\FFF_i$ individually, but the expressions are very lengthy and not very informative. Instead, we proceed with calculating the hybridization matrix $\DD(z)= [\GG(z)]^{-1} \sum_i \FFF_i \VV_i$.  We furthermore assume that $t$ is real and introduce the dimensionless quantity $\tt = \pi \rho t$. We also set $\phi_{\rm L}=\phi/2$ and $\phi_{\rm R}=-\phi/2$.
Finally, we take the large-$\Delta$ limit of the lead propagator

\beq{
\GG_{ik} 
= \frac{1}{z^2-(\Delta_i^2+\epsilon_k^2)} \begin{pmatrix}
z+\epsilon_k & e^{i\phi_1}\Delta_i & 0 & 0 \\
e^{-i\phi_i} \Delta_i & z-\epsilon_k & 0 & 0 \\
0 & 0 & z+\epsilon_k & -e^{i\phi_i}\Delta_i \\
0 & 0 & -e^{-i\phi_i}\Delta_i & z-\epsilon_k
\end{pmatrix}
}
to obtain
\beq{
\GG_i = \sum_k \GG_{ik} = 
-\pi \rho \begin{pmatrix}
0 & e^{i\phi_i} & 0 & 0 \\
e^{-i\phi_i}  & 0 & 0 & 0 \\
0 & 0 & 0 & -e^{i\phi_i} \\
0 & 0 & -e^{-i\phi_i} & 0
\end{pmatrix}.
}
With these assumptions and simplifications in place, the hybridisation matrix can be written as
\beq{
\DD(z) = 
\frac{\pi \rho}{1+2\tt^2\cos\phi+\tt^4}
\begin{pmatrix}
-2a & 
b & -2 c & 0 
\\
b^* & 2 a & 0 & 2c
\\
-2c & 0 & -2 a & 
-b
\\
0 & 2c & -b^*
& 2a 
\end{pmatrix},
}
where we introduced the following notation
\beq{
\begin{split}
a &= (V_{\rm L} V_{\rm R}+W_{\rm L} W_{\rm R})\tt(\tt^2+\cos\phi), \\
b_i &= V_i^2 + W_i^2, \\
b &= e^{-i\phi/2}(b_{\rm R}+\tt^2 b_{\rm L}) + e^{+i\phi/2}(b_{\rm L} + b_{\rm R} \tt^2), \\
c &= (V_{\rm R} W_{\rm L} - V_{\rm L} W_{\rm R}) \tt\sin\phi.
\end{split}
}
The key feature of this expression is that this matrix includes terms in the
out-of-diagonal $2\times2$ blocks. These correspond to the presence of an effective magnetic field in the $x$ direction that induces the spin polarization along this same direction. 
The particular direction ($x$) results from the assumed form of the spin-orbit-coupling terms in Eq.~\eqref{H} and from assuming real $V_{\rm L, R}$, $W_{\rm L, R}$ and $t$.
In terms of second quantization operators, the hybridisation matrix corresponds to the following form:
\beq{
\Delta_\mathrm{hyb} = \frac{\pi \rho}{1+2\tt^2\cos\phi +\tt^4}
\left( -2a \sum_\sigma d^\dag_\sigma d_\sigma - 4c S_x + b
d^\dag_\uparrow d^\dag_\downarrow + b^* d_\downarrow d_\uparrow +
\text{const.} \right).
}
Not all terms contribute in both spin sectors. The pairing terms proportional to $b$ are only relevant in the spin-singlet sector, while the effective spin-splitting terms proportional to $c$ are only relevant in the spin-doublet sector. The potential term proportional to $a$ contributes in both subspaces. Since we are interested only in the $S=1/2$ subspace, in the following we concentrate on this particular $2\times2$ subspace. We find that the effective Hamiltonian of the doublet sector in the ${\uparrow,\downarrow}$-basis is given by
\begin{equation}\label{eq:Heff}
H_{\rm eff} = \begin{pmatrix} 
\epsilon + E_{z}/2 & \left(E_{x} -i E_{y}\right)/2 \\
\left(E_{x} + i E_{y}\right)/2 & \epsilon - E_{z}/2
\end{pmatrix} - \frac{2\pi \rho}{1+2\tilde{t}^2\cos{\phi}+\tilde{t}^4} \begin{pmatrix} 
a & c \\
c & a 
\end{pmatrix}.
\end{equation}
This model is exact in the double limit $U_{ee} \to 0$, $\Delta_{\rm L, R} \to \infty$.
In general, one expects correction factors to parameters that depend on both $\Delta_{\rm L, R}$ and $U_{ee}$, which control the energy cost of charge fluctuations from the doublet state.
These corrections can be accurately computed using the NRG method. 
Nevertheless, the general form remains the same, as confirmed by numerical calculations, see Sec.~\ref{Sss:NRG}.
Most importantly, the conditions for the matrix element $c$ to be non-zero, as revealed in this calculation, hold fully generally, and are the following: a) the presence of additional QD levels (i.e., nonzero parameter $t$ in the generalized SIAM), b) the presence of both spin-preserving and spin-flip tunneling (so that the combination $V_{\rm R} W_{\rm L}-V_{\rm L} W_{\rm R}$ is non-zero, which is expected to be generally true except in cases of accidental cancellation), c) finite phase bias $\phi$.

We need to note that in the superconducting atomic limit the doublet state in the standard superconducting-SIAM model does not depend on the phase bias $\phi$, as can be checked by taking the limit $W_{\rm L, R} \to 0$ and $t\to 0$ in Eq.~\eqref{eq:Heff}.
However, away from the superconducting atomic limit an additional diagonal term $E_D \cos(\phi)$ arises, with $E_D>0$. This term is generated by fourth-order processing in hopping (second order in hybridisation) and has a minimum at $\phi=\pi$~\cite{spivak1991}, as typical for Josephson junctions with an odd-parity ground state.

Assuming $V_{\rm L}=V_{\rm R} \equiv V$, $W_{\rm L} = -W_{\rm R} \equiv W$ (note that this sign for $W_i$ choice merely reflects the sign convention in the Hamiltonian and actually corresponds to the symmetric situation with the same amplitude for the left SC to QD and for the QD to right SC spin-flip tunneling), and defining
\begin{equation*}
\Gamma_V = \pi \rho V^2, \quad \Gamma_W = \pi \rho W^2,
\end{equation*} 
the second term of Eq.~\eqref{eq:Heff} can be written as 
\begin{equation}
\frac{2\tilde{t}}{1+2\tilde{t}^2\cos{\phi}+\tilde{t}^4} \begin{pmatrix} 
\left(\Gamma_V-\Gamma_W\right)\left(\tilde{t}^2+\cos{\phi}\right) & 2\sqrt{\Gamma_V\Gamma_W}  \sin{\phi} \\
2\sqrt{\Gamma_V\Gamma_W}  \sin{\phi} & \left(\Gamma_V-\Gamma_W\right)\left(\tilde{t}^2+\cos{\phi}\right)
\end{pmatrix}.
\label{eq:H_effectiveWVt}
\end{equation}
Assuming that $\tilde{t}\ll 1$, we can simplify the model further by performing a series expansion to obtain
\begin{equation}
2 \tilde{t} \begin{pmatrix} 
\left(\Gamma_V-\Gamma_W\right)\cos{\phi} & 2\sqrt{\Gamma_V\Gamma_W}  \sin{\phi} \\
2\sqrt{\Gamma_V\Gamma_W}  \sin{\phi} & \left(\Gamma_V-\Gamma_W\right)\cos{\phi}
\end{pmatrix}.
\end{equation}
Defining 
\begin{equation}
\label{eq:ESO}
E_{t} = 2\tilde{t} \left(\Gamma_V-\Gamma_W\right),
\quad
E_{\rm SO} = 4\tilde{t} \sqrt{\Gamma_V\Gamma_W}\,,
\end{equation}
we find the approximate small $\tilde{t}$ Hamiltonian given by (up to a constant): \begin{equation}
\label{seq:2x2ham}
H_{\rm eff} = \begin{pmatrix} 
E_{z}/2 & \left(E_{x} -i E_{y}\right)/2 \\
\left(E_{x} + i E_{y}\right)/2 & - E_{z}/2
\end{pmatrix} - \begin{pmatrix} 
E_{t} \cos{\phi} & E_{\rm SO} \sin{\phi} \\
E_{\rm SO} \sin{\phi} & E_{t} \cos{\phi} 
\end{pmatrix}.
\end{equation}
This expression takes the form of the phenomenological potential for the transmon circuit given by main text Eq.~(1).
Given that there is a potential cancellation of the $E_t$ term, it is prudent to include in the model an addition term of the form $E_D \cos(\phi)$ from processes that are higher-order in hybridisation.
This term will combine with $-E_t \cos(\phi)$ to produce the $+E_0 
\cos(\phi)$ potential with $E_0=E_D-E_t$ in Eq.~(1).
Note that $E_0$ can take either a positive or negative sign.

Within the limit considered here, we find that $E_{\rm SO}$ depends on each of the three types of coupling in the model: spin-conserving, spin-flipping, and direct lead-lead tunneling. All three have to be present for the spin-splitting to occur. Furthermore, it may happen that the cosine term drops out if the prefactors of all contributions add up to zero,
resulting in a Josephson potential shifted by $\pi/2$ compared to the singlet state. This fine-tuned situation is indeed encountered in the experiment as discussed in Sec.~\ref{Sss:differentSO}.

It is instructive to evaluate the eigenvalues of the isolated quantum dot junction. In the simplified model of Eq.~(1) these are given by
\begin{equation}
E_{\uparrow,\downarrow} = E_{0} \cos{\phi} \pm \frac{1}{2}\sqrt{E_{y}^2+E_{z}^2 + \left(E_{x} - 2 E_{\rm SO} \sin{\phi}\right)^2}.
\end{equation}
For $E_{y} = E_{z} = 0$, this simplifies to 
\begin{equation}
E_{\uparrow,\downarrow} = E_{0} \cos{\phi} \pm \left(E_{x}/2 - E_{\rm SO} \sin{\phi}\right)
\end{equation}
The Zeeman field parallel to $E_{\rm SO}$ enters as a constant offset, which does not change the curvature of the potential and does not affect the transmon frequency. Furthermore, this results in the spin-flip transition frequency given by
\begin{equation}
E_{\uparrow}-E_{\downarrow} = E_{x} - 2E_{\rm SO} \sin{\phi}
\end{equation}
which is linear in the applied Zeeman field. Setting $E_{x} = E_{y} = 0$ instead, we find 
\begin{equation}
E_{\uparrow,\downarrow} = E_{0} \cos{\phi} \pm \frac{1}{2}\sqrt{E_{z}^2 + 4 E_{\rm SO}^2 \sin^2{\phi}}
\end{equation}
Here the $E_{z}$ term does enter the curvature of the potential, thus affecting the transmon frequency. Furthermore, the resulting in the spin-flip qubit transition frequency is given by
\begin{equation}
E_{\uparrow}-E_{\downarrow} = \sqrt{E_{z}^2 + 4 E_{\rm SO}^2 \sin^2{\phi}}
\end{equation}
The presence of the $\sin^2{\phi}$ term results in the doubling in periodicity we observe in the perpendicular field dependence of main text Fig.~4(c) compared to the parallel field dependence of Fig.~4(b).

\subsection{NRG calculations}\label{Sss:NRG}

The proposed model has very little symmetry: the spin-orbit coupling fully breaks the rotational SU(2) spin symmetry, and the BCS mean-field approximation breaks the U(1) charge conservation. The only remaining symmetry is Z$_2$ fermionic number parity (even or odd total number of electrons in the system). Furthermore, the Hamiltonian has complex-valued matrix elements. Nevertheless, the quantum impurity problem can still be solved using the conventional impurity solver, the numerical renormalization group (NRG), albeit at quite significant computational cost. The NRG method consists of discretizing the continua (two superconducting baths), their transformation into Wilson tight-binding chains, and an iterative diagonalization of the resulting chain/ladder Hamiltonian. We performed a very coarse discretization with the discretization parameter $\Lambda=8$; nonetheless, for the purposes of computing energy splitting, this remains a surprisingly good approximation. We retain up to 3000 states in each NRG step. On 8 cores of an AMD EPYC 7452 processor, such NRG calculations take approximately 15 minutes for each parameter set. The band is assumed to have a constant density of states in the interval $[-D:D]$. In the following, all model parameters will be given in units of half-bandwidth $D$.

\begin{figure}[t!]
    \centering
        \includegraphics[scale=1.0]{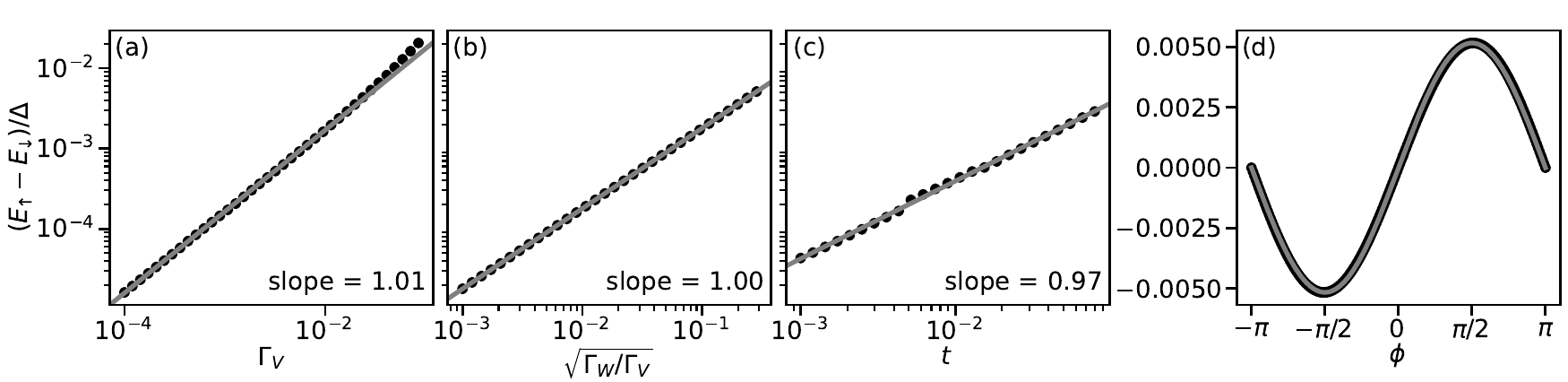}   
\caption{Scaling of the induced spin splitting in the doublet state with parameters $\Gamma_V$, $\Gamma_W$, and $t$, as well as the $\phi$-dependence. Unless stated otherwise,
the parameter set is $U_{ee}=1$, $\epsilon=-U_{ee}/2$, $\Gamma_V=0.02$, $\Gamma_W=\Gamma_V/5$, $t=0.1$, $\Delta=0.01$, $\phi=\pi/4$.
(a) $\Gamma_V$-dependence, demonstrating the linearity of splitting as a function of the hybridisation strength, with small non-linear corrections for large $\Gamma_V$.
(b) $\Gamma_W/\Gamma_V$-dependence, demonstrating the linearity of splitting as a function of the ratio of spin-flip over spin-preserving tunneling processes to the impurity orbital in resonance.
(c) $t$-dependence, demonstrating the linearity of splitting as a function of cotunneling through non-resonant impurity levels.
(d) $\phi$-dependence, showing a very clean $\sin(\phi)$ behavior of the spin-splitting, as predicted by the reduced analytical model. 
 }
    \label{fig:scaling1}
\end{figure}

We first verify the findings from Supplementary Sec.~\ref{Sss:analytics}, specifically the spin splitting induced by the combination of the spin-flip scattering, the presence of multiple levels in the QD (represented by the interdot tunneling term $t$), and finite superconducting phase difference between the two SC contacts, as described by the $E_\mathrm{SO}\sin\phi$ terms in the effective Hamilotinian with $E_\mathrm{SO}=4\tt \sqrt{\Gamma_V \Gamma_W} = 4 \tt \Gamma_V \sqrt{\Gamma_W/\Gamma_V}$, see Eqs.~\eqref{eq:ESO} and \eqref{eq:ESO}. In Fig.~\ref{fig:scaling1} we plot the dependence of the splitting $E_\uparrow-E_\downarrow$ as a function of key parameters. We indeed observe that the splitting is linear in the hybridisation $\Gamma$ for fixed $\Gamma_W/\Gamma_V$ ratio, in the ratio $W/V =\sqrt{\Gamma_W/\Gamma_V}$ and in the hopping $t$. Finally, we also ascertain the $\sin(\phi)$ dependence of the splitting. We have thus confirmed that the splitting is linear in $t$, $\Gamma$, $W/V$, and $\phi$ for small parameter values.

\begin{figure}[t!]
    \centering
        \includegraphics[scale=1.0]{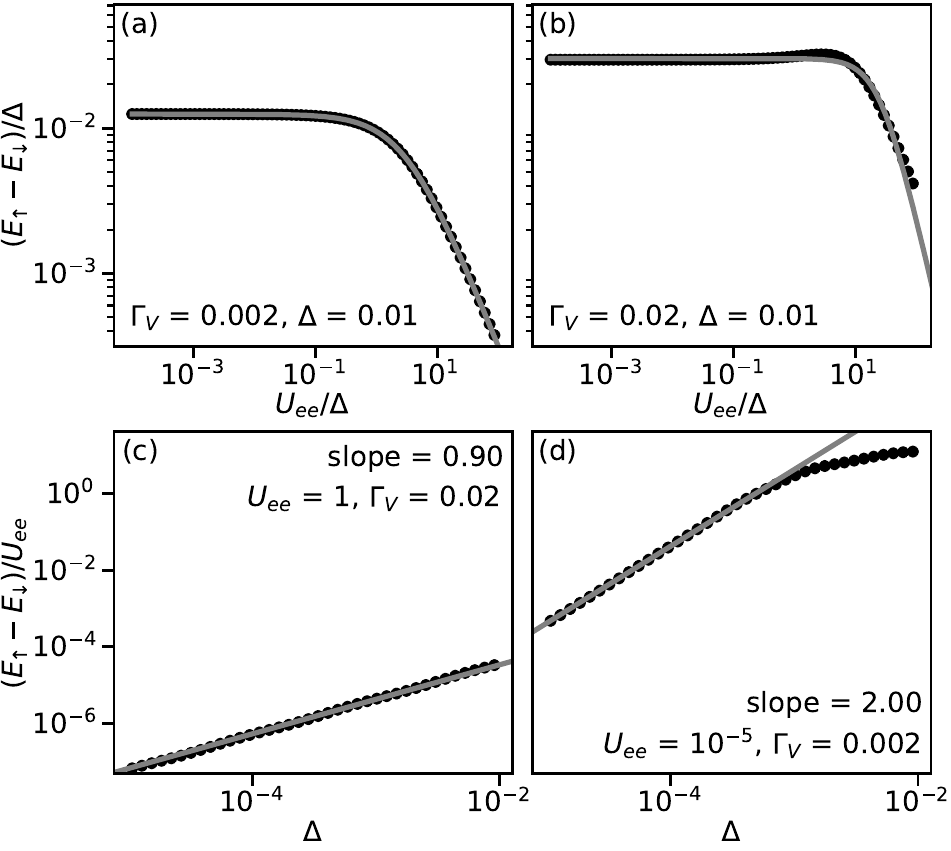}   
\caption{ Scaling of the induced spin splitting in the doublet state with parameters $U_{ee}$ and $\Delta$. Other parameters as in Fig.~\ref{fig:scaling1}.
(a) $U_{ee}$-dependence at $\Gamma_V=0.002$ showing the cross-over from the $U_{ee}\ll\Delta$ regime to the $U_{ee} \gg \Delta$ regime.
(b) Same as a, but for stronger hybridisation $\Gamma_V=0.02$, showing the more complex behavior away from the low-$\Gamma_V$ limit.
(c) $\Delta$-dependence at $\Gamma_V=0.02$, showing that the splitting is roughly proportional to $\Delta$ in
the $U_{ee}\gg\Gamma,\Delta$ regime.
(d) Same as (c), but for $\Gamma_V=0.002$ and much smaller $U_{ee}=10^{-5}$ (non-interacting limit), showing quadratic scaling for $\Delta \ll \Gamma$ that crosses over into linear scaling for $\Delta \gg \Gamma$.
 }
    \label{fig:scaling2}
\end{figure}

The dependence on other model parameters, in particular $U_{ee}$ and $\Delta$, is not simple. The parameters $U_{ee}$ and $\Delta$ control the energy cost of charge fluctuations, and the behavior depends not only on their ratio, but also on their values compared to the hybridisation $\Gamma$ as well as the bandwidth $D$. The simplest case is the linear regime of small parameter $\Gamma$, where the splitting is simply inversely proportional to $1/(U_{ee}/2+\Delta)$ to a good
approximation, see Fig.~\ref{fig:scaling2}(a). For larger $\Gamma$, we observe deviations from this simple form, see Fig.~\ref{fig:scaling2}(b). It is also instructive to consider the dependence on $\Delta$ at fixed $U_{ee}$. The limit of small $U_{ee}$ is merely of academic interest, because the doublet state is then a (highly) excited state: we find a roughly linear dependence on $\Delta$, see Fig.~\ref{fig:scaling2}(c).  For large $U_{ee}$, however, we find a complex dependence that furthermore depends on the value of $\Gamma$, showing a cross-over from
quadratic dependence for $\Delta \ll \Gamma$ to a roughly linear dependence for $\Delta \gtrsim \Gamma$, see Fig.~\ref{fig:scaling2}(d). From these plots we conclude that the dependence on $U_{ee}$, $\Delta$ and $\Gamma$ (when $\Gamma$ is not small) is highly non-trivial and should in general be computed numerically (e.g. using the NRG method).



\begin{figure}[t!]
    \centering
        \includegraphics[scale=1.0]{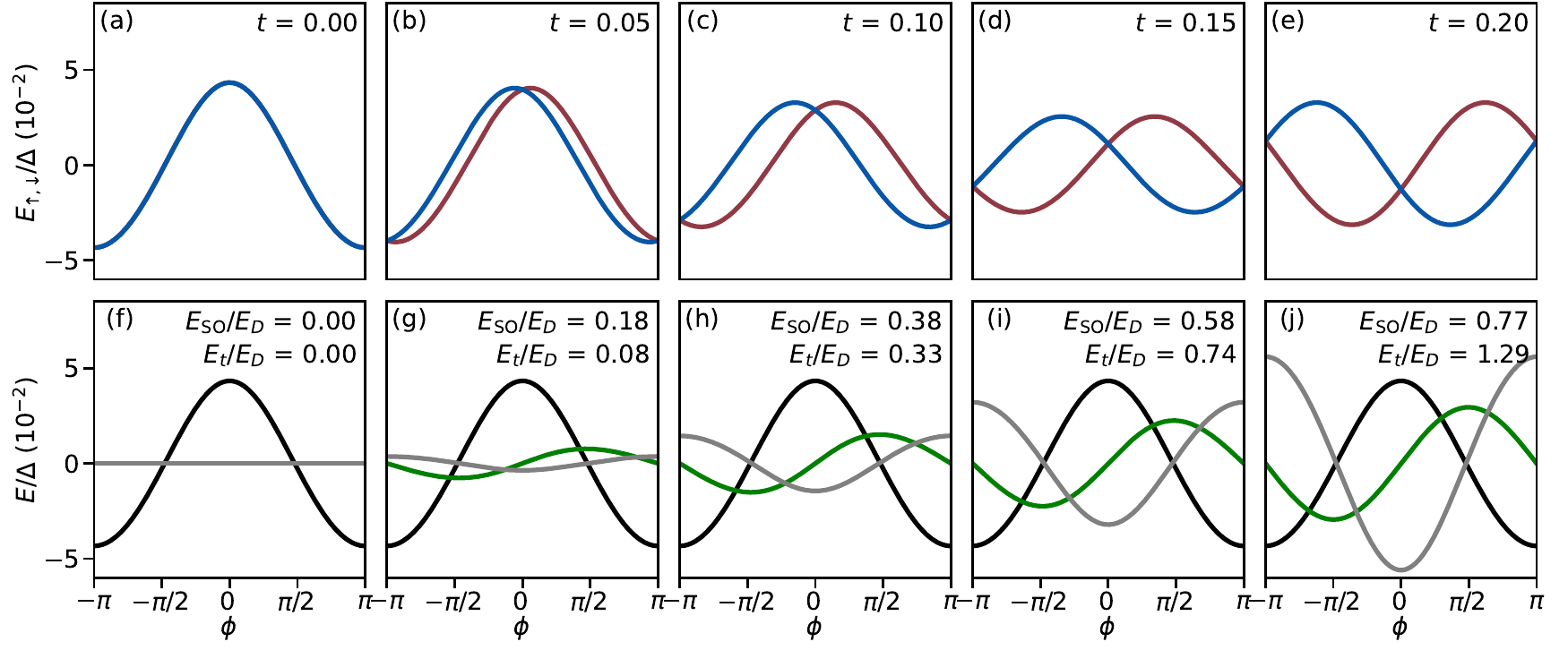}   
\caption{ Decomposition of the doublet potential energy into its components. We show the results for various values of $t$, as indicated for each panel.
All plots have the same axis ranges in order to permit direct comparison of magnitudes. (a-e) Total potential for the two spin states. (f-g) $E_D \cos(\phi)$ potential, common to all cases (black), $-E_t \cos(\phi)$ (green) and $E_\mathrm{SO} \sin(\phi)$ (grey) contributions. We find $E_D=4.33\, 10^{-3}$ (same for all $t$) and the 
$E_t/E_D$ and $E_\mathrm{SO}/E_D$  ratios indicated for each panel.
Other model parameters are $U_{ee}=1$, $\epsilon=-U_{ee}/2$, $\Delta=0.1$, $\Gamma_V=0.2$, $\Gamma_W/\Gamma_V=1/5$.
}
    \label{fig:sincos}
\end{figure}

In Fig.~\ref{fig:sincos} we explore the three contributions to the doublet potential: the conventional doublet $E_D \cos(\phi)$ potential with the minima at $\phi=\pm \pi$, the $E_\mathrm{SO} \sin(\phi)$ potential due to spin-flip scattering with minima at $\phi=\pm \pi/2$, as well as the $-E_t \cos(\phi)$ potential due to cotunneling though the multiple levels of the QD with minimum at $\phi=0$, see Fig.~\ref{fig:sincos}. We plot the results for a range of $t$ starting from zero; this case serves as a reference from which we extract the standard $E_D$ part. With increasing $t$, both $E_\mathrm{SO}$ as well as $E_t$ increase. This displaces the minima in the effective potential from $\phi=\pm \pi$ towards $\phi=\pm \pi/2$. When $E_t$ becomes equal to $E_D$, the $\cos(\phi)$ part of the potential cancels out. For $E_t > E_D$, the minima move past $\phi = \pm \pi/2$ and tend toward $\phi=0$.


\newcommand{\SSS}{\tilde{S}}

\begin{figure}[t!]
    \centering
        \includegraphics[scale=1]{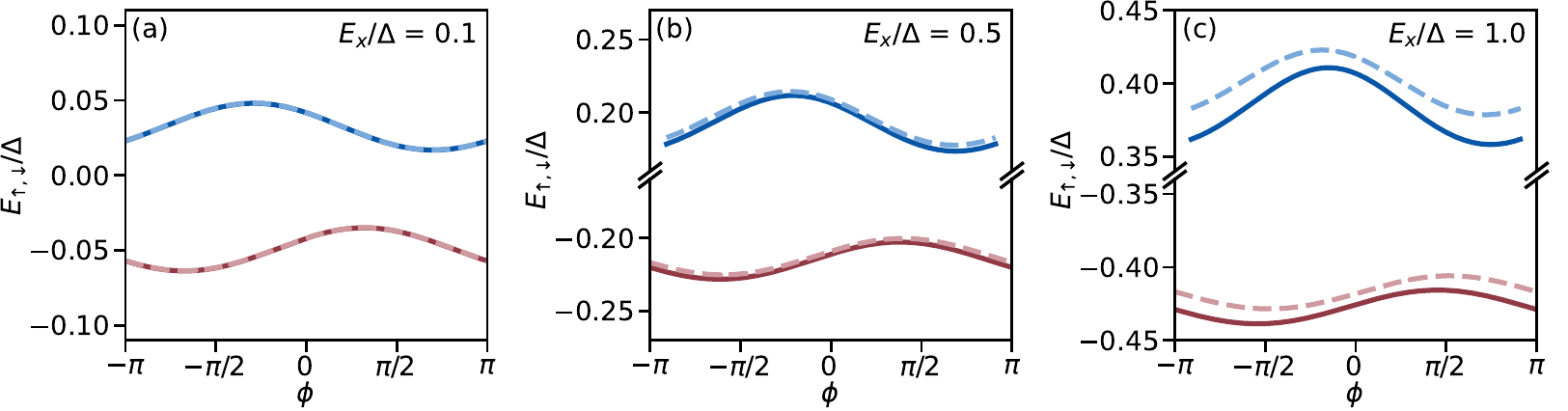}    
        \caption{ Comparison of the energies of the doublet subgap states as a function of the phase bias computed using
        the standard NRG procedure (solid lines) and using the ``marginal-operator trick'' approximation (discontinuous lines). Vertical scale is the same for all panels. Model parameters are $U_{ee}=1.5$, $\Gamma_{\rm L}=\Gamma_{\rm R} \equiv \Gamma = 0.2$, $\gamma_{\rm L}=\gamma_{\rm R}=0.2\Gamma$, $t=0.1$, $\epsilon=0.1-U_{ee}/2$, $\Delta=0.1$. }
    \label{fig:motnrg}
\end{figure}

A major time-saving procedure is to incorporate the effects of the external magnetic field as a perturbation to the results of an NRG calculation for a Hamiltonian without any  field terms. This ploy rests on the observation that the impurity spin operators are exactly marginal (in the renormalization-group sense): their matrix elements remain of the same order of magnitude throughout the NRG iteration, i.e., they neither blow up nor decay to zero. The method may hence be dubbed the ``marginal-operator trick''. The idea is to perform the NRG iteration of spin operators $S_x$, $S_y$, $S_z$ through unitary transformations to find the effective spin operators in the NRG eigenbasis in the low-energy sector. These are then added to the effective Hamiltonian with bare Zeeman energies $E_x$, $E_y$, $E_z$:
\beq{
H_\mathrm{eff} = \sum_w E(w) \ket{w} \bra{w} + E_x \SSS_x + E_y \SSS_y + E_z \SSS_z.
}
Here $w$ indexes the eigenstates $\ket{w}$ with eigenenergies $E(w)$, while $\SSS_i$ are the transformed spin matrices in this same basis. The basis can be truncated to a small number of levels; in many cases it is sufficient to retain solely the subgap states. This effective Hamiltonian may then be diagonalized at negligible numerical cost for arbitrary values of $E_x$, $E_y$ and $E_z$. In case where only two (subgap) states are retained one can even write down closed-form expressions for eigenenergies and eigenstates. The marginal-operator trick is a good approximation up to Zeeman energies comparable to the BCS energy gap $\Delta$, as it has been ascertained by comparisons with the NRG calculations with the Zeeman terms included from the outset, see Fig.~\ref{fig:motnrg}. This method is clearly very generally applicable to any problem involving marginal operators in the Hamiltonian, obviating the need for costly parameter sweeps in multidimensional spaces.

\subsection{Transmon diagonalization}
\label{Sss:transmon}
Having established how to calculate an effective potential in the doublet sector, we now turn to its inclusion in the Hamiltonian of the encompassing transmon circuit \cite{Koch2007}:
\begin{equation}
H = -4E_{\rm c}\partial_\phi^2 + E_{\rm J}(1-\cos\delta) + U(\phi),
\label{eq:transmon_ham}
\end{equation}     
where $E_{\rm c}$ and $E_{\rm J}$ denote the charging energy of the transmon island and the Josephson energy of the reference junction, respectively, and $U(\phi)$ denotes the effective doublet potential of main text Eq.~(1).
The two phase drops across the quantum dot junction ($\phi$) and across the reference junction ($\delta$) are connected according to $\phi-\delta=\phi_\textrm{ext}$, where $\phi_{\rm ext}=(2e/\hbar)\Phi_\textrm{ext}$ is the phase difference resulting from the externally applied magnetic flux through the SQUID loop, $\Phi_\textrm{ext}$ [Fig.~1(d)].

Following \cite{Bargerbos2020, Kringhoj2020b}, we numerically diagonalize \eqref{eq:transmon_ham} in the phase basis. This results in the energy levels $E_n$ as well as the associated transition frequencies $f_{nm} = \left(E_m - E_n\right)/h$, capturing both the transmon and the spin-flip transitions. Having calculated the transition frequencies, fits can be made to the data. This is done to obtain the estimates for the effective model parameters found in the main text and in the next sections, using $E_{\rm c}/h=\SI{284}{MHz}$ and $E_{\rm J}/h=\SIrange{12.4}{12.7}{GHz}$ as reference junction parameters. Note that the reference junction gate voltage is generally held fixed in the experiment, and that the range in $E_{\rm J}$ is the result of cross-coupling between the quantum dot and reference junction gate lines.

Here we note that the sinusoidal reference junction potential used in Eq.~\eqref{eq:transmon_ham} is that of a conventional superconductor-insulator-superconductor (SIS) tunnel junction, governed by many weakly transparent channels. Previous work has found that nanowire-based Josephson junctions are more accurately described by several or even a single transport channel, leading to a more skewed potential shape \cite{Kringhoj2018}. This can result in a reduction of the the qubit anharmonicity, and thus an underestimation of $E_{\rm{c}}$ when using the SIS potential. However, the inclusion of a more involved potential introduces additional fitting parameters, and obtaining unique solutions is not guaranteed. This holds in particular because the reference junction is operated far from its pinchoff voltage, such that several channels are expected to contribute to the potential (see Sec.~\ref{Sss:tuneup}). We therefore choose to use the SIS potential throughout the Letter. In practice, this choice affects the value of $E_{\rm{c}}$ that is extracted from the fit, which in turn rescales the extracted values of $E_0$ and $E_{\rm SO}$. 

\clearpage

\section{Device and experimental setup}
\label{Ss:device}

\subsection{Device overview}
\begin{figure}[h!]
    \center
        \includegraphics[scale=1.0]{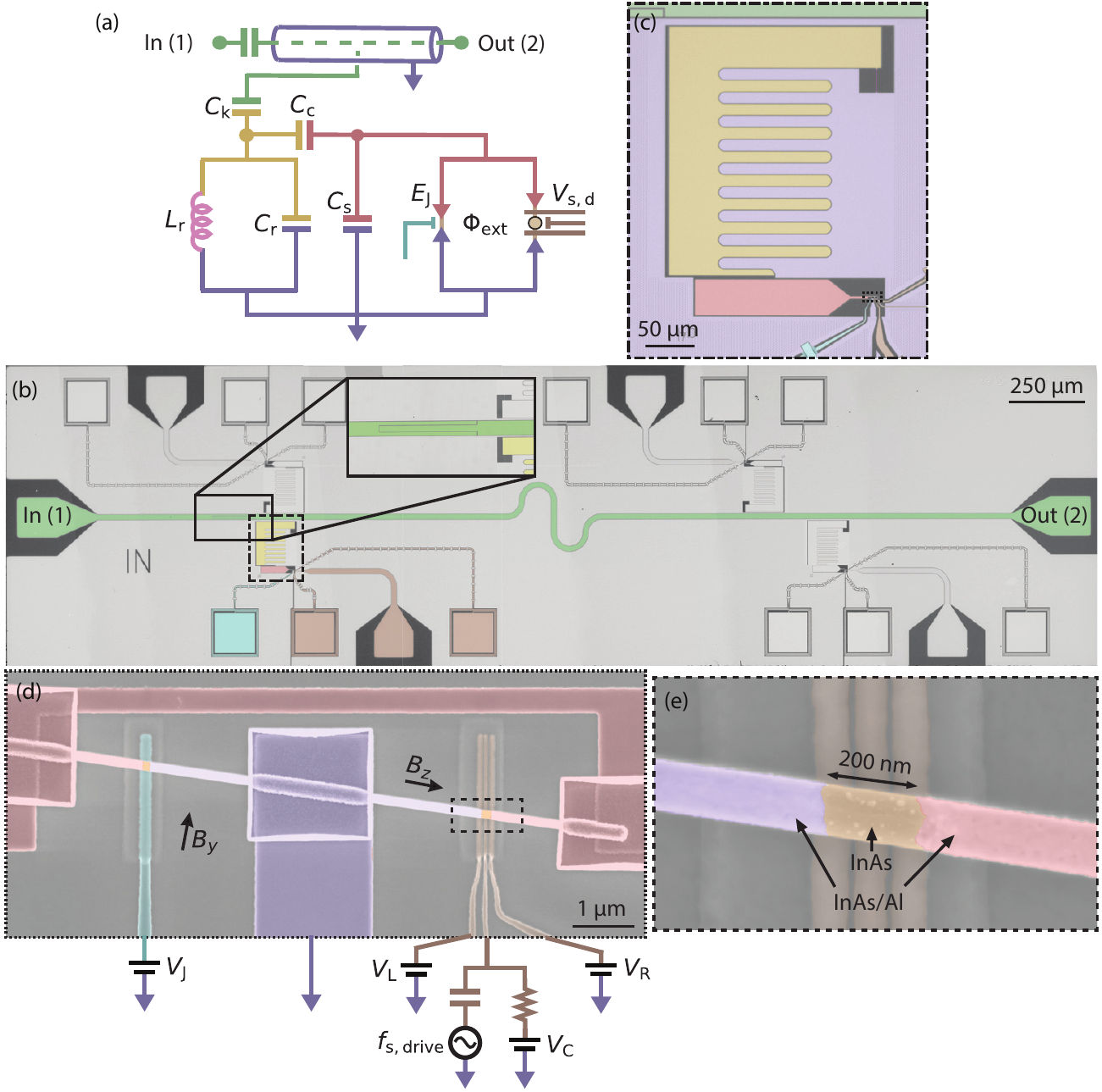}    
        \caption{{ \bf Device overview.}  (a) Diagram of the microwave circuit. A coplanar waveguide transmission line with an input capacitor (green center conductor) is capacitively coupled to a grounded LC resonator. The resonator consists of an island (yellow) capacitively and inductively  (pink) shunted to ground (blue). The resonator is in turn capacitively coupled to a transmon island (red), which is shunted to ground capacitively as well as via two parallel Josephson junctions.     
        (b) Chip containing four nearly identical devices coupled to the same transmission line, which has an input capacitor, enlarged in inset. 
        (c) False-colored optical microscope image of the device showing the qubit island, the resonator island, the resonator inductor, the transmission line, the electrostatic gates and ground. 
        (d) False-colored scanning electron micrograph (SEM) of a nearly identical device, showing the InAs/Al nanowire into which the junctions are defined. The $B_y$ component of the magnetic field is used to tune $\Phi_{\rm ext}$ \cite{Wesdorp2021}. $B_z$ is the magnetic field component parallel to the nanowire. 
        (e) False-colored SEM of a nearly identical device, showing the quantum dot junction in which the quantum dot is gate defined. The three bottom gates have a width and spacing of \SI{40}{nm}, although this is obfuscated by the dielectric layer placed on top.
        }
        
    \label{fig:device}
\end{figure}

The physical implementation of the device studied is shown in Fig.~\ref{fig:device}. It is analogous to that of \cite{Bargerbos2022}, repeated here for convenience.
The chip, 7~mm long and 2~mm wide, consists of four devices coupled to a single transmission line with an input capacitor to increase the directionality of the outgoing signal  [Fig.~\ref{fig:device}(b)].
For the experiments performed in this Letter only two of the devices were wire-bonded: the device measured in the main text, and a second device, which was not functional.

For each device, a lumped element readout resonator is capacitively coupled to the feedline [Fig.~\ref{fig:device}(c)]. The resonator is additionally capacitively coupled to the transmon island, which is connected to ground via a SQUID loop formed by the reference and quantum dot junctions [Fig.~\ref{fig:device}(d)]. Both junctions are implemented on a single \SI{10}{um}-long epitaxial superconductor-semiconductor nanowire with a \SI{100}{nm}-wide hexagonal InAs core and a \SI{6}{nm}-thick Al shell covering two of its facets \cite{Krogstrup2015}. They are defined in two uncovered nanowire sections (\SI{110}{nm}-long for the reference junction and \SI{200}{nm}-long for the quantum dot junction). A zoom-in of the the quantum dot junction is shown in Fig.~\ref{fig:device}(e). The reference junction is controlled by a single \SI{110}{nm}-wide electrostatic gate, set at a DC voltage \Vj. The quantum dot junction is defined by three \SI{40}{nm}-wide gates separated from each other by \SI{40}{nm}. We note that in Fig.~1(e) of the main text the gates appear wider (and the gaps between gates appear smaller) than stated due to distortion by the gate dielectric layer. The outer two gates are set at DC voltages \Vl~and \Vr. The central gate is connected to a bias-tee  formed by a \SI{100}{\kilo\ohm} resistor and a \SI{100}{pF} capacitor. This permits the simultaneous application of a DC signal \Vc~ to control the level of the quantum dot junction and a microwave tone $f_{\rm s, drive}$ to drive the spin-flip transition. 

\subsection{Nanofabrication details}

The device fabrication occurs in several steps using standard nanofabrication techniques. It is identical to that described in~\cite{Bargerbos2022}, and repeated here for the sake of completeness. The substrate consists of 525~$\upmu$m-thick high-resistivity silicon, covered in \SI{100}{nm} of low pressure chemical vapor deposited $\rm{Si_3N_4}$. On top of this, a \SI{20}{nm} thick NbTiN film is sputtered, into which the gate electrodes and circuit elements are patterned using an electron-beam lithography mask and $\rm{SF_6}$/$\rm{O_2}$ reactive ion etching. Subsequently, \SI{30}{nm} of $\rm{Si_3N_4}$ dielectric is deposited on top of the gate electrodes using plasma enhanced chemical vapor deposition and then etched with a buffered oxide etchant. The nanowire is then deterministically placed on top of the dielectric using a nanomanipulator and an optical microscope. After placement, two sections of the aluminium shell are selectively removed by wet etching with MF-321 developer. These sections form the quantum dot junction and the reference junction, with lengths \SI{200}{nm} and \SI{110}{nm} respectively. After the junction etch, the nanowire is contacted to the transmon island and to ground by an argon milling step followed by the deposition of \SI{150}{nm}-thick sputtered NbTiN. Finally, the chip is diced into 2 by 7 millimeters, glued onto a solid copper block with silver epoxy, and connected to a custom-made printed circuit board using aluminium wirebonds. 

\clearpage

\subsection{Cryogenic and room temperature measurement setup}

\begin{figure}[h!]
    \center
    \includegraphics[scale=0.6]{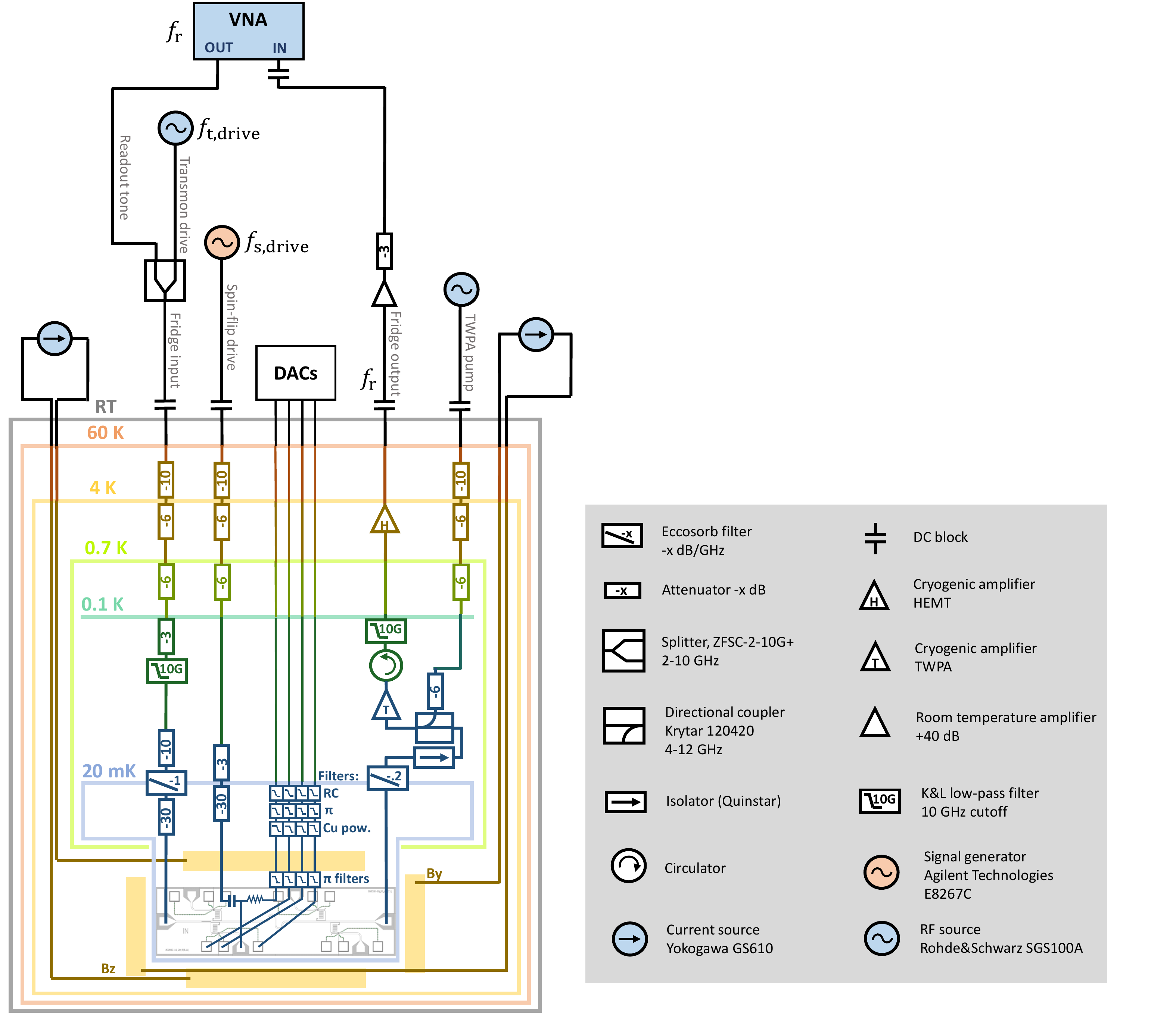}
    \caption{{\bf Measurement setup at cryogenic and room temperatures.} The device was measured in a Triton dilution refrigerator with a base temperature of \SI{20}{mK}. The setup contains an input RF line, an output RF line, an additional RF line for driving the spin-flip transition, and multiple DC gate lines. The DC gate lines are filtered at base temperature with multiple low-pass filters connected in series. The input and drive RF lines contain attenuators and low-pass filters at different temperature stages, as indicated. The output RF line contains a travelling wave parametric amplifier (TWPA) at the \SI{20}{mK} temperature stage,  a high-electron-mobility transistor (HEMT) amplifier at the \SI{4}{K} stage, and an additional amplifier at room temperature. A three-axis vector magnet (x-axis not shown) is thermally anchored to the \SI{4}{K} temperature stage, with the device under study mounted at its center. The three magnet coils are  controlled with Yokogawa GS610 current sources. At room temperature, a vector network analyzer (VNA) is connected to the input and output RF lines for spectroscopy at frequency $f_{\rm r}$. On the input line, this signal is then combined with the transmon drive tone at frequency $f_{\rm t, drive}$, for two-tone spectroscopy. The spin-flip drive tone at frequency $f_{\rm s, drive}$ is sent through the additional RF line.}
    \label{fig:cryogenic_setup}
\end{figure}

\section{Basic characterization and tune up}
\label{Sss:tuneup}
This section describes how the device is tuned to its setpoints. 

\begin{figure}[h!]
    \center
    \includegraphics[scale=1.0]{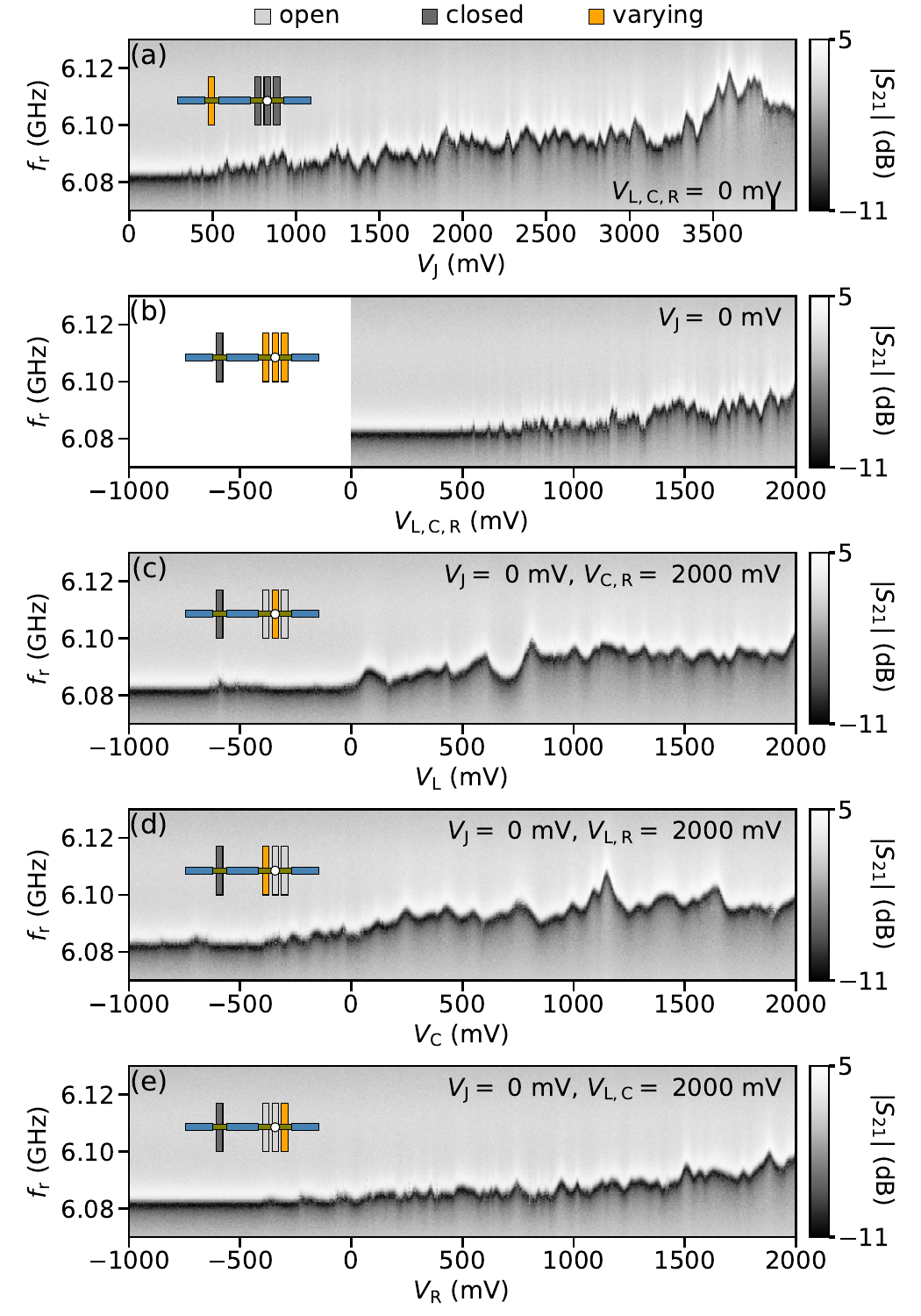}    
    \caption{{ \bf Electrostatic gates characterization.}  Transmission amplitude as a function of frequency of a single tone $f_{\rm r}$ and gate voltage. For each panel, the inset indicates which gate is being varied (orange) and which ones are set to a value above (light grey) or below (dark grey) their pinch-off value.}
    \label{fig:pinchoffs}
\end{figure}

We start by characterizing the effect of the electrostatic gates, which control each of the two Josephson junctions. Fig.~\ref{fig:pinchoffs}(a) shows the basic behaviour of the reference junction versus junction gate voltage \Vj~when the quantum dot junction is completely closed. As \Vj~is varied, different junction channels open sequentially \cite{Spanton2017, Hart2019}, with transparencies that increase non-monotonically due to mesoscopic fluctuations. This in turn affects the $E_{\rm J}$ of the transmon, allowing for in-situ tunability of its frequency, and the transmon then affects the resonator through its dispersive shift \cite{Blais2004}, resulting in the observed change in resonator frequency. We use this to choose a \Vj~set-point which maintains a good SQUID asymmetry in all regimes of interest. The black line in Fig.~\ref{fig:pinchoffs}(a) indicates \Vj~=~\SI{3860}{mV}, the setpoint used in Figs.~3(a-b) in the main text. After a small non-reproducible gate jump in the reference junction the setpoint was retuned to \Vj~=~\SI{4064.5}{mV}, which was used for all other data shown in the main text. For all resonances explored, we maintained $E_{\rm J}/E_{\rm c} > 40$.

In Figs.~\ref{fig:pinchoffs}(b-e) we show analogous measurements where we vary the quantum dot gate voltages when the reference junction is closed. We first measure an effective pinch-off curve for all three quantum dot gates ramped together [Fig.~\ref{fig:pinchoffs}(b)], before sweeping each gate separately,  with the other two quantum dot gates kept at \SI{2000}{mV} [Figs.~\ref{fig:pinchoffs}(c-e)]. This shows that each of the three quantum dot gates can independently pinch off the quantum dot junction,  even if the other gates are in the open regime, signifying strong lever arms and good gate alignment. Note that these are not pinch-off curves as encountered in conventional tunnel spectroscopy; they reflect the voltages at which there is no longer a measurable transmon transition frequency mediated by the quantum dot junction, which could either be due to low tunneling rates or a full depletion of the quantum dot. 

The subsequent tuning procedure for finding an isolated quantum dot resonance is discussed in detail in Ref.~\cite{Bargerbos2022}, summarized here for the specific resonances used in the main text. First we close the reference junction and go to a point in quantum dot gate voltages near pinchoff.  Fixing the readout frequency $f_{\rm r}$ at the bare frequency of the resonator, one can then map out the regions where dispersive shifts occur on a two-dimensional map versus the left and right quantum dot gates, with the central gate held fixed. This signifies regions in which there is a supercurrent flowing through the quantum dot junction. After identifying such a region in \Vl-\Vr~space, we subsequently open the reference junction, which lifts the reference transmon frequency closer to the bare resonator frequency. This magnifies the dispersive shift of the resonator and, furthermore, brings the external flux into the picture. Fixing \flux~=~0 and repeating the initial measurement then reveals much stronger deviations of the resonant frequency due to the enhanced dispersive shift. 

\begin{figure}[h!]
    \center
    \includegraphics[scale=1.0]{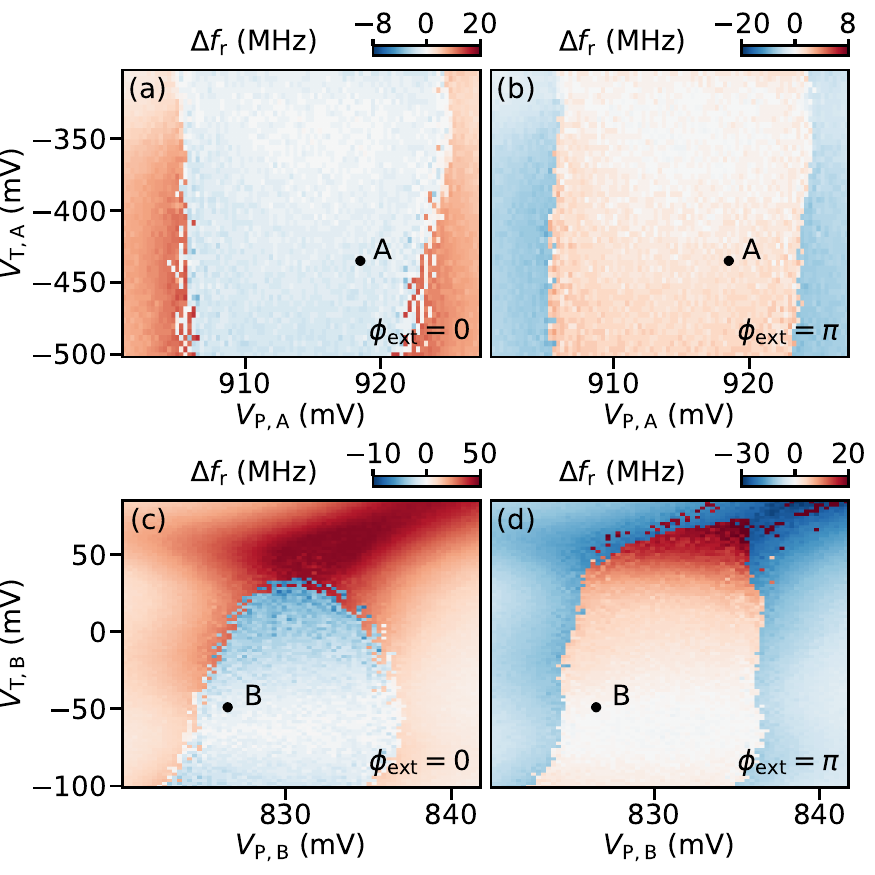}    
    \caption{(a,c) Shift of the resonator resonance frequency with respect to its value when the quantum dot junction is fully closed, $\Delta f_{\rm r}$, versus \Vp~and \Vt~ at \flux~=~0, revealing singlet (red) and doublet (blue) ground state regions separated by sharp transitions for resonance A (a) and resonance B (c). (b,d) Same as (a,c) but for \flux~=~$\pi$.} 
    \label{fig:resonances}
\end{figure}

Using this approach we identify isolated quantum dot resonances, and subsequently explore their evolution versus the central quantum dot gate. This is shown in Fig.~\ref{fig:resonances}, where we furthermore account for cross coupling between the different quantum dot gates by defining a new set of virtual gates. For simplicity we fix \Vl~and focus on the rotated \Vr-\Vc~space, denoted as the \Vp-\Vt-space. Note that this compensation scheme is unique for each isolated region we explore. Fixing \flux~=~0 and varying the central dot gate, the resonator first shows a displacement towards higher frequencies to then abruptly drop to a lower frequency, to then finally go back to the higher frequencies once-more [Fig.~\ref{fig:resonances}(a,c)]. This behaviour is reversed for $\phi_{\rm ext} = \pi$ [Fig.~\ref{fig:resonances}(b,d)], and can be identified as a singlet-doublet transition resulting from the relative level of the quantum dot is being varied by \Vp~ \cite{Bargerbos2022}. We note that in the \Vt~direction we do not always find the expected dome shape characteristic of singlet-doublet transitions; while such a shape does develop for resonance B [Fig.~\ref{fig:resonances}(c-d)], the doublet phase of resonance A [Fig.~\ref{fig:resonances}(a-b)] remains open even at elevated tunnel gate voltages. This is potentially a result of a non-monotonic dependence of tunnel rates on the gate voltage.

\clearpage

\section{Extended data}
\subsection{Spin-orbit splitting at different resonances} 
\label{Sss:differentSO}
\begin{figure}[h!]
    \center
    \includegraphics[scale=1.0]{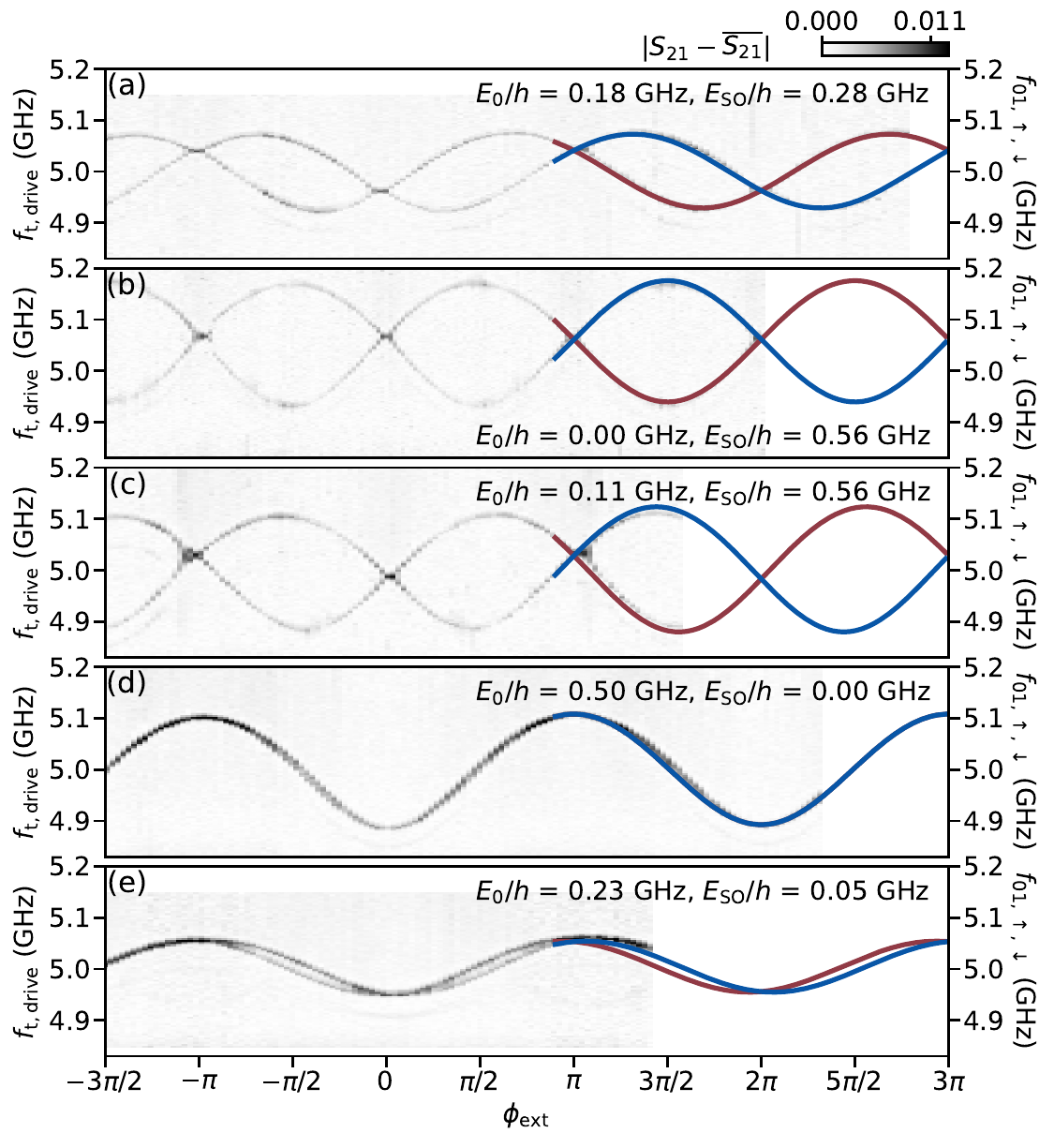}    
    \caption{{ \bf Spin-splitting energies at different resonances.}  Flux dependence of transmon spectroscopy taken at different points in quantum dot gate space, indicated in Table~\ref{tab:gates}. In all cases, the quantum dot junction is in a doublet state. Different panels show different spin-splitting energies.
    (a) and (b) are  gate setpoints A and B of the main text, respectively. 
    }
    \label{fig:splittings}
\end{figure}

As discussed in the main text, we find a wide variety of phase-dependent splittings depending on the quantum dot resonance studied. This is shown in Fig.~\ref{fig:splittings}, portraying a range of resonances all the way from an even phase dependence with no splitting (panel d) to resonances that have a fully odd phase dependence (panel b). By Fitting the potentials with the transmon Hamiltonian, as dicussed in Sec.~\ref{Sss:transmon}, we extract a set of effective parameters for each resonance, tabulated in Table~\ref{tab:gates}. 

\begin{table*}[h!]
\caption{
Quantum dot junction gate voltage set points and extracted model parameters for the five panels in Fig.~\ref{fig:splittings}.
}
\label{tab:gates}
\begin{ruledtabular}
\begin{tabular}{ccccccc}
\textrm{}& $V_{\rm L}$ (mV)  & $V_{\rm C}$ (mV)  & $V_{\rm R}$ (mV) & $E_{0}/h$ (GHz) & $E_{\rm SO}/h$ (GHz) \\
\colrule
(a)\footnote{Setpoint A in the main text figures.} & 79.0 & 1020.0 & 363.0 & 0.18 & 0.28 \\
(b)\footnote{Setpoint B in the main text figures.} & 430.0 & 531.0 & 635.2 & 0.00 & 0.56  \\
(c) & 430.0 & 520.0 & 652.0 & 0.11 & 0.56\\
(d) & 430.0 & 614.8 & 335.0 & 0.50 & 0.00 \\
(e) & 430.0 & 655.2 & 305.2 & 0.23 & 0.05 \\
\end{tabular}
\end{ruledtabular}
\end{table*}


\subsection{Spin-orbit splitting within the same resonance}
\label{Ss:2DmapESO}

\begin{figure}[h!]
    \center
    \includegraphics[scale=1.0]{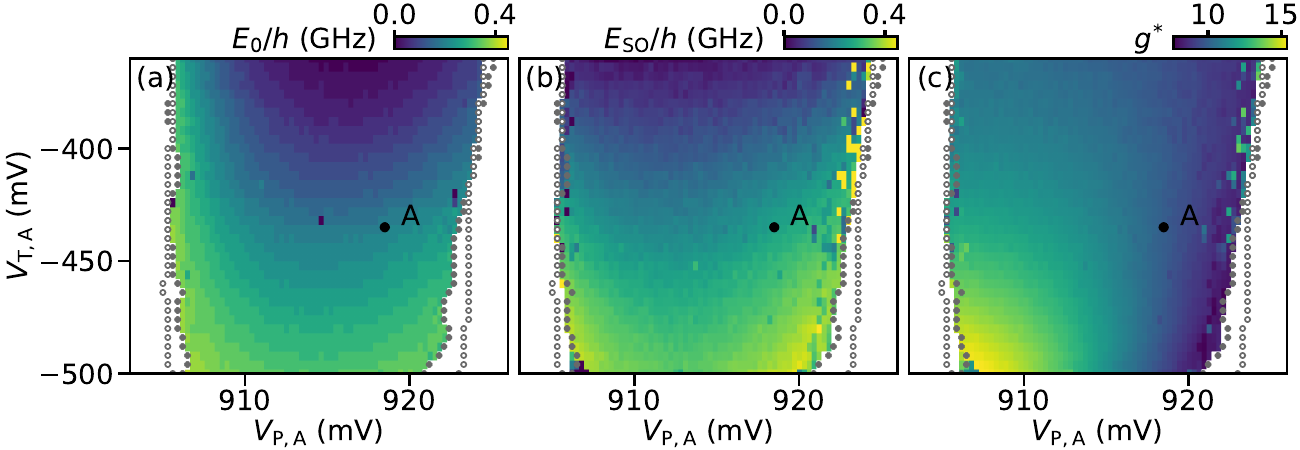}    
    \caption{{ \bf Gate dependence within resonance A.} (a) Magnitude of $E_0$ extracted from transmon qubit spectroscopy at $\phi_{\rm ext} = 0, \pi$ versus rotated plunger and tunnel gate voltages, at a magnetic field of \SI{18}{mT} applied parallel to the nanowire. (b) Same as (a) for $E_{\rm SO}$, extracted from spin-flip spectroscopy at $\phi_{\rm ext} = \pi/2, 3\pi/2$. (c) Same as (b) for $g^{*}$, extracted from the same measurement as (b). 
    }
    \label{fig:gatemap}
\end{figure}

Within the extended SIAM model, $E_{\rm SO}$ and $E_{0}$ are expected to depend on $\Gamma_V$, $\Gamma_W$, and $\tilde{t}$ (see Sec.~\ref{Ss:theory}). One would therefore expect that these quantities can also vary within a single resonance, as the gate voltages tune the relative energy levels as well as the tunnel barriers of the quantum dot. This is indeed observed in the experiment: as shown in Fig.~\ref{fig:gatemap}(a,b) for resonance A, we find that both effective doublet parameters vary with the rotated plunger and tunnel gates. In particular, both $E_{0}$ and $E_{\rm SO}$ show an increase towards the boundary of the singlet doublet transition, i.e. towards the edges of the Coulomb diamond. This is in line with the predictions of Ref.~\cite{Padurariu2010}, as in the middle of the Coulomb diamond the energy cost of adding an electron to the quantum dot is maximal and the high energy of the intermediate states reduces the probability of Cooper pair tunneling. Additionally, contrary to initial expectation, the magnitude of the effective parameters appears to decrease for larger tunnel gate values. While the tunnel gate is expected to increase the tunnel rates, and thus the effective parameters, we note that in practice the situation is highly complex; there are up to three gate voltages that control five model parameters ($\Gamma_{V,W}^{\rm L,R}$, $\tilde{t}$), with potentially non-monotonic dependencies as well as cross-coupling. A full understanding of such a system will require a more detailed study of such dependencies, which we leave for future work. At this stage we instead emphasize the gate-tunability of these quantities, allowing for in-situ fine-tuning of the model parameters 

Furthermore, we also find that the effective Landé g-factor $g^{*}$ depends on gate voltage [Fig.~\ref{fig:gatemap}(c)], in line with previous results on quantum dots in InAs nanowires, demonstrating its electric gate tunability \cite{Csonka2008}. This could be of relevance for qubit applications, as the tunability can be used to rapidly drive spin states in and out of resonance with a static magnetic field induced electron spin resonance condition. Finally, we note that the observed gate dependence of $g^{*}$ is distinct from that of $E_{\rm SO}$ and $E_{0}$, supporting the assertion that its origin is tied to a complex interplay of spin-orbit coupling and confinement, beyond the model considered here \cite{Kiselev1998, Schroer2011, Winkler2017}.

\subsection{Magnetic field angle dependence and determination of the spin-splitting direction} \label{Sss:angle}

\begin{figure}[ht!]
    \center
    \includegraphics[scale=1.0]{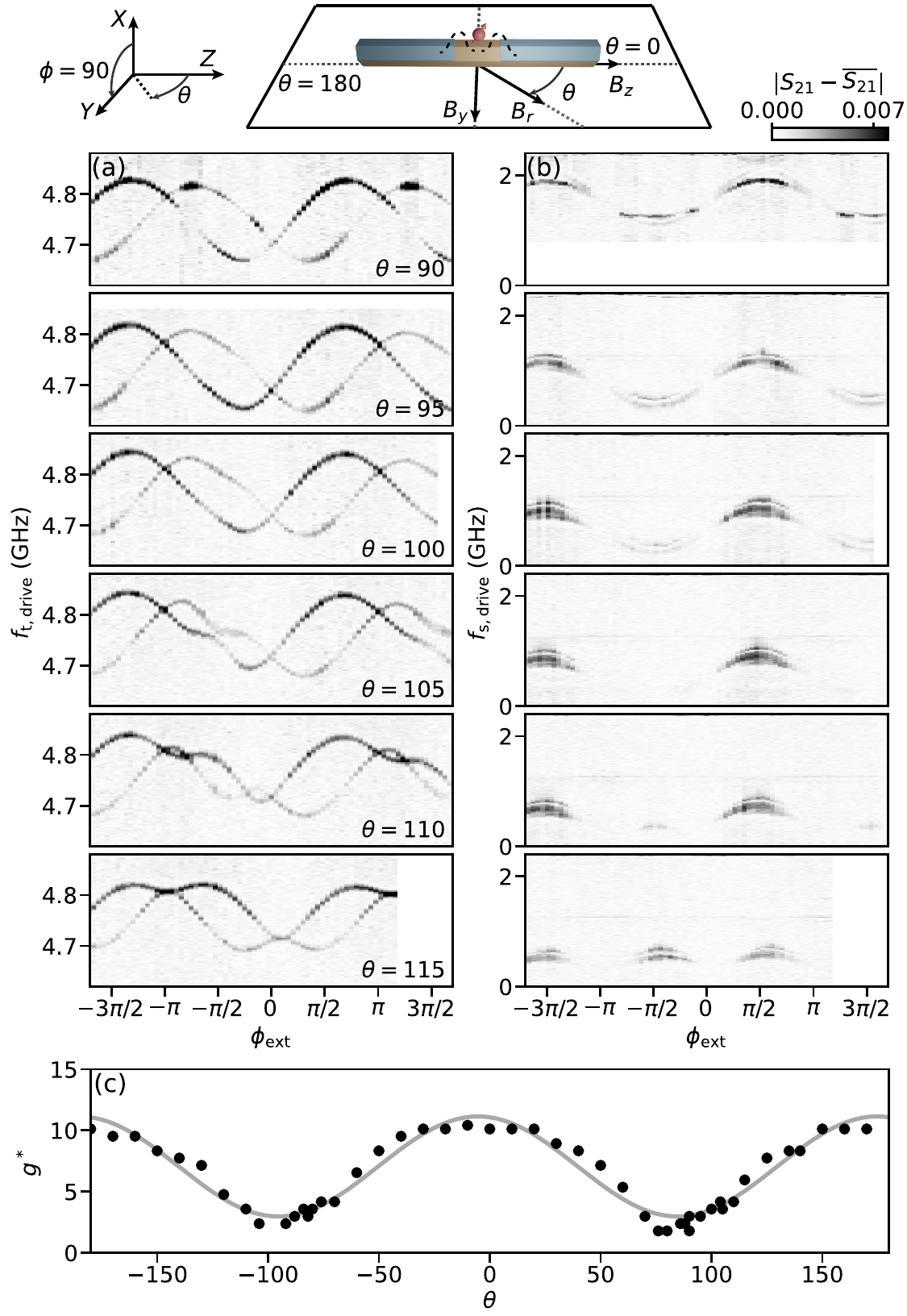}    
    \caption{{ \bf Magnetic field angle dependence of resonance A.} (a-b) Flux dependence of transmon (left column) and spin-flip (right column) transitions for different magnetic field directions, for a fixed total magnetic field $B_r=$~\SI{12}{mT}. Each row corresponds to a different magnetic field orientation on the chip plane, $\phi=90$,  determined by the angle $\theta$ with respect to the nanowire direction (see diagram at the top). All panels share the same color bar. (c) Effective Landé g-factor $g^*$ versus $\theta$. Markers show data extracted from spin-flip spectroscopy at \SI{12}{mT}, and the solid line shows a fit with a cosine.
    }
    \label{fig:theta}
\end{figure}

In this Section we detail the method used to determine the direction of the spin-orbit interaction at a fixed gate point (vector $\vec{n}$ in Eq.~(1) in the main text).
This is done by comparing the angle dependence of transmon and spin-flip spectroscopy to the predictions of the model discussed in Sec.~\ref{Ss:theory}.
For this it is useful to define a coordinate space determined by the nanowire direction, $Z$, the on-chip direction perpendicular to the nanowire, $Y$, and the direction perpendicular to the chip, $X$.
We then define $\theta \in [0,180)$ as the polar angle with respect to the $Z$~direction and $\phi \in [0, 360)$ as the azimuthal angle [see Fig.~\ref{fig:theta}], such that $x = B_r \cos(\phi) \sin(\theta)$, $y = B_r \sin(\phi) \sin(\theta)$ and $z = B_r \cos(\theta)$.
With this convention, $\phi=90$~is the plane of the chip, while $\phi=0$ is the plane perpendicular to the chip containing the nanowire.
In each of these two planes, we first fix the magnitude of the applied magnetic field, $B_r$, and measure the evolution of transmon and spin-flip spectroscopy while varying $\theta$ in steps of two degrees.
For each plane we determine the angle for which the applied field is perpendicular to the spin-splitting direction, $\theta_{\perp, 0}$ and $\theta_{\perp, 90}$, by comparison to the theory model. The cross product of these two directions determines the direction parallel to the spin-splitting term. 

Representative data of such a measurement for resonance A is shown in Fig.~\ref{fig:theta}(a-b), where we fix $\phi=90$ and find $\theta_{\perp, 90}=78$. Performing an analogous measurement in the $\phi=0$ plane, we determine $\theta_{\perp, 0}=86$.
From these two, we obtain $(\theta_{\rm s}, \phi_{\rm s}) = (167, 72)$ as the spin-splitting direction $\vec{n}$ of resonance A.
This is 13 degrees away from the nanowire axis. Furthermore, we generally find that the measured spin-split direction varies depending on which quantum dot resonance is studied; for resonance B of the main text we obtain a spin-splitting direction of $(\theta_{\rm s}, \phi_{\rm s}) = (72, 45)$, which is 84 degrees away from the spin-splitting direction of resonance A.

We can furthermore estimate the effective Landé g-factor from the evolution of the spin-flip transition frequency versus the angle of the magnetic field. Shown in Fig.~\ref{fig:theta}(c) for resonance A, the effective g-factor varies from 3 to 11 depending on the angle, minimal for magnetic fields perpendicular to the spin-orbit direction.
The measured behaviour is well-described by a simple cosine, in line with previous results on quantum dots in InAs nanowires \cite{Schroer2011}.

\subsection{Spin-flip spectroscopy enabled by spin-orbit splitting}
As discussed in the main text, we do not rely on driving  transitions of the transmon circuit to perform spectroscopy of the junction's excitation spectrum. While in principle possible by using three microwave tones, this could result in limitations due to the finite transmon lifetime as well as undesired mixing processes between the different tones. Instead, we use the dispersive shift from the transmon's ground state to induce a doublet-state-dependent shift on the resonator, similar to how the island parity of offset-charge sensitive transmon qubits can be distinguished \cite{Serniak2019, Uilhoorn2021}. As the difference between the transmon frequencies of both doublet states is small, inducing a sizeable dispersive shift larger than the resonator's linewidth requires us to tune the spin-dependent transmon qubit frequency close to that of the resonator [Fig.~\ref{fig:device}(a-b)]. Having done so, we can observe the spin-flip transition directly with conventional two-tone spectroscopy, where the first tone is applied at the frequency of the readout resonator, and the second tone at the spin-flip frequency [Fig.~\ref{fig:device}(c)]. The transmon, off-resonant from both tones, remains in its ground state during the measurement.

\begin{figure}[ht!]
    \center
    \includegraphics[scale=1.0]{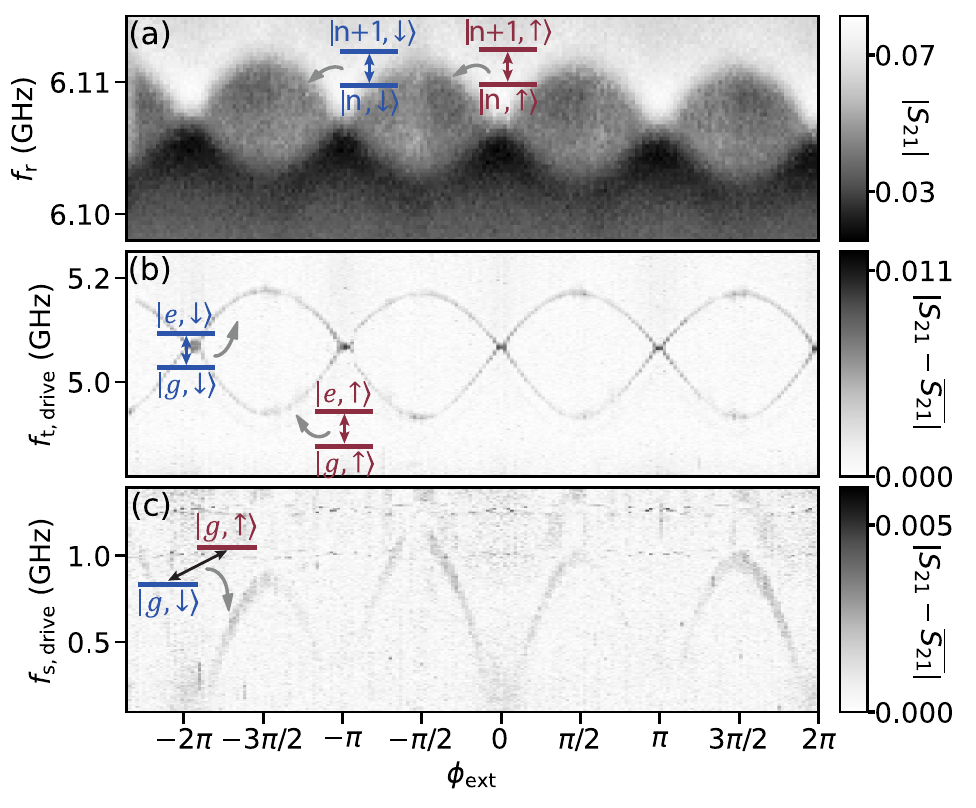}    
    \caption{{ \bf Spin-flip two-tone spectroscopy.} (a) Flux dependence of single-tone spectroscopy showing the resonator frequency. Each of the two visible branches corresponds to a different spin state of the quantum dot junction. (b)  Flux dependence of two-tone spectroscopy showing the transmon frequency. Each of the two visible branches corresponds to a different spin state of the quantum dot junction. (c)  Flux dependence of two-tone spectroscopy showing the spin-flip frequency. 
    }
    \label{fig:spinflip}
\end{figure}

\bibliography{supplement.bib}